\documentclass[journal]{IEEEtran}

\makeatletter

\newcommand{\Rmnum}[1]{\expandafter\@slowromancap\romannumeral #1@}
\makeatother

\usepackage[numbers,sort&compress]{natbib}
\usepackage{url}
\usepackage{bm}
\usepackage{amsfonts}
\usepackage{amsmath}
\usepackage{amssymb}
\usepackage{amsthm}
\usepackage{color}
\usepackage{enumerate}
\usepackage{makecell}

\usepackage{array,color}%使用表格
\usepackage{tabularx}%使用表格
\usepackage{multirow}%使用多栏宏包
\usepackage{booktabs}%定义了三条划线命令：\toprule、\midrule 和 \bottomrule，可分别对表格顶部、中部和底部使用不同粗细的水平线
\usepackage{makecell}  %table

\usepackage{graphicx}  %figure
\usepackage{subfig}
\usepackage{epstopdf}
\usepackage{graphics}
\usepackage{epsfig}
\usepackage{psfrag}
\usepackage{overpic}

\usepackage{multicol}

\usepackage{stfloats}
\usepackage{float}   %强制H

\usepackage{indentfirst} %缩进
\usepackage{gensymb}

\usepackage{color, colortbl}

\begin{document}

\title{A Hybrid Framework for Topology Identification of Distribution Grid with Renewables Integration}
\author{Xing~He,~\IEEEmembership{Member,~IEEE},  Robert~C. Qiu,~\IEEEmembership{Fellow,~IEEE}, Qian~Ai,~\IEEEmembership{Senior Member,~IEEE},  Tianyi~Zhu
\thanks{This work was partly supported by  National Key Research \& Development (R\&D) plan of China (grant No. 2016YFB0901300), and National Natural Science Foundation of China (grant No. 51907121 and No. U1866206).}
%\thanks{Xing~He, Qian~Ai, Robert~C. Qiu are with the Department of Electrical Engineering, Research Center for Big Data Engineering Technology, State Energy Smart Grid R$\&$D Center, Shanghai Jiaotong University, Shanghai 200240, China. (e-mail: {hexing\_hx@126.com)}}% <-this % stops a space
%%National Natural Science Foundation of China
}
\maketitle
\begin{abstract}
Topology identification (TI) is a key task for state estimation (SE) in distribution grids, especially the one with high-penetration renewables.
The uncertainties, initiated by the time-series behavior of renewables, will \textit{almost certainly}  lead to bad TI results \textit{without a proper treatment}.
These  uncertainties are \textit{analytically intractable} under conventional framework---they are usually \textit{jointly spatial-temporal dependent}, and hence cannot be simply treated as white noise.
For this purpose, a hybrid framework is suggested in this paper to handle these uncertainties in a \textit{systematic and theoretical} way; in particular, big data analytics are studied \textit{to harness the jointly spatial-temporal statistical properties} of those uncertainties.
With some prior knowledge, a model bank is built first to store the \textit{countable} typical models of network configurations;
therefore, the difference between the SE outputs of each bank model and our observation \textit{is capable of being defined as a matrix variate}---the so-called \textit{random matrix}.
In order to gain insight into the random matrix, a \textit{well-designed metric space} is needed.
Auto-regression (AR) model, factor analysis (FA), and random matrix theory (RMT) {are tied together for} the metric space design, followed by \textit{jointly temporal-spatial analysis} of those matrices which is conducted \textit{in a high-dimensional (vector) space}.
Under the proposed framework, some \textit{big data analytics} and \textit{theoretical results} are obtained to improve the TI performance.
Our framework is validated using IEEE standard distribution network with some field data in practice.
\end{abstract}


\IEEEpeerreviewmaketitle
\section{Introduction}
\label{Introd}
\IEEEPARstart{T}opology identification (TI) of admittance matrix $\mathbf {Y}$, the so-called network topology, is a precondition for state estimation (SE) in distribution systems.
Inaccurate TI has long been cited as a major cause of bad SE results \cite{Wu1989Detection}.
During a daily operation, $\mathbf {Y}$  may be partially reconfigured \cite{bolognani2013identification}.
While the knowledge of $\mathbf {Y}$ is crucial, it may be unavailable or outdated (via TI) due to some reasons~\cite{cavraro2018graph, deka2017structure, ardakanian2019identification, yu2017patopa, yuan2016inverse}. Among these reasons, the uncertainties caused by the behavior of  high-penetration renewables  \cite{he2019invisible, yang2018robust}, which are \textbf{analytically intractable} for most tools, are one of the main challenges.  \textbf{How to address these uncertainties by harnessing their jointly spatial-temporal statistical properties} is at the heart of our study, and this question threads throughout the proposed hybrid framework.

%The $\mathbf {Y}$ may not be accurately obtained for the following reasons: 1) the uncertainty, diversity, and individuality of system units, especially those renewables which are small in size but large in amount \cite{he2019invisible}; 2) the status of the switches, most of which are maintained manually, may be out-of-data due to erroneous record, delay telemetry, or unexpected operation; 3) line impedance parameters are susceptible to climate changes and usage lifetime; 4) inevitable measurement errors, e.g., data missing, abnormality, or out of sync; and 5) ubiquitous noise, e.g., load/DG fluctuations \cite{yang2018robust}.

\subsection{Related Work and Motivation of our Work}

Ref.~\cite{singh2010recursive, cavraro2017power, tian2015mixed} are {relevant to our paper to an extent}. 
Ref.~\cite{singh2010recursive} builds a {model bank}, and then conducts TI task by applying a recursive Bayesian approach to identify the correct network configuration in the bank.  Ref.~\cite{cavraro2017power} conducts TI task by comparing the collected {voltage time series} with a {library of signatures computed a priori}.
Ref.~\cite{tian2015mixed}  formulates the TI problem as a mixed integer quadratic programming (MIQP) model to find a topology configuration with weighted {least square} (WLS) of {measurement residues}.

Several {data-driven} TI approaches, as Ref.~\cite{cavraro2018graph, deka2017structure, ardakanian2019identification, yu2017patopa, yuan2016inverse}, are proposed recently.
They are mainly based on iterations, graph theory, sparsity-based regularization,  and so on.
These approaches are feasible to TI task {with very little knowledge} about the network.
Ref.~\cite{yu2017patopa} tells that an accurate TI result is acquirable \textbf{only if the noise is well addressed}. For instance, even with a small error in measurements, the regression-based method {may fail in TI task} (see {Sec.~\ref{sec:RegandF}} in  \textit{Case Studies}).
Most TI algorithms, especially those derived from least square, rely heavily on the {second-order statistics} of meter data~\cite{deka2017structure, ardakanian2019identification}, and hence they are applicable to (Gaussian) white noise.

Renewables-derived uncertainties (e.g., randomness caused by a gust of wind), however, often \textbf{exhibit themselves as non-Gaussian noise}.
The conventional statistics such as first/second-order statistics (mean/variance) {are even not nearly enough} to represent these non-Gaussian variables, and the \textbf{(jointly spatial-temporal) dependence should be taken into account}.
Therefore, there is an urgent need for some powerful approach to make these uncertainties {analytically tractable} with \textbf{a systematic and theoretical procedure}.
This is the \textbf{{major motivation and superiority}} of our proposed hybrid framework.
Under our framework, some \textbf{statistical properties and theoretical results} are established.

\subsection{Our Work and its Contributions}

In order to handle the renewables-derived uncertainties, we \textbf{have to} go back to the model bank following Ref.~\cite{singh2010recursive, cavraro2017power}. It is \textbf{reasonable and feasible to list all the possible models} in practice with prior knowledge, since the network configuration of a particular grid {must be confined to only a few typical models}.
Because of the bank,  the {difference} between the bank model SE output and our observation {is capable of being defined} as a matrix variate---the so-called \textbf{{random matrix}}.

Then we move to the heart of our hybrid framework---{high-dimensional analytics}  of the random matrix.
Auto-regression (AR) model, factor analysis (FA), and random matrix theory (RMT) are tied together for the \textbf{jointly temporal-spatial modeling and analysis} of the random matrices.
And {high-dimensional statistics} are obtained as big data analytics.
This framework {enables us} to gain insight into the (multiple) renewables-derived uncertainties, which are analytically intractable under conventional framework.

In particular, our framework deals with a large number (spatial space, $N$) of nodes simultaneously, and each node ($i\!=\!1,..., N$) samples {time-series} within a given duration (temporal space, $T$) of observation. Classical statistic theories treat  fixed $N$ only (often small, typically $N\!<\!6$ \cite{qiu2015smart}) . This fixed (small) $N$ is called the {low-dimensional regime}. In practice, we are interested in the case that $N$ can {vary arbitrarily in size compared with} $T$ (often $T$ is large, {typically} $N\!>\!20$, $c\!=N/T\!>0$ \cite{qiu2015smart}). This fundamental requirement is the  {primary driving force} for us to study big data analytics with {high-dimensional statistics}.
For jointly spatial-temporal analysis, \textbf{a (large-dimensional) data matrix}, rather than a vector or a scalar \cite{7364281}, is adopted {as the basis}.

This work is expected to contribute some insight to the (multiple) renewables-derived uncertainties that are often analytically intractable.
We take advantage of high-dimensional statistics that is made {analytically tractable only recently}~\cite{yeo2016random,ding2019spiked}. To our knowledge, this type of analysis is, \textbf{for the first time}, conducted in the context of TI.
Our big data analytics are motivated to improve TI performance, and may be further expanded to other applying fields: the detection and localization of faults \cite{he2011dependency}, the detection of unmonitored switching of circuit breakers in network reconfiguration~\cite{sharon2012topology}, etc.

The remainder of this paper is organized as follows.
\begin{enumerate}[-]
\item Sec.~\ref{Sec:Framework} presents the hybrid framework and gives a general discussion about it.
\item Sec.~\ref{Sec:ProFor}, by employing the model bank, aims to convert our observation into a random matrix with prior knowledge.
\item Sec.~\ref{Sec:Resi} studies the high-dimensional statistics of the random matrices based on AR, FA, and RMT.
\item Sec.~\ref{Sec:Cases} validates our framework with case studies based on IEEE standard distribution network using some field data.
\end{enumerate}

\section{Hybrid Framework  of Topology Identification}
\label{Sec:Framework}
\subsection{Hybrid Framework}
\label{Sec:Assumpt}
\begin{figure*}[htbp]
\centering
\includegraphics[width=0.92\textwidth]{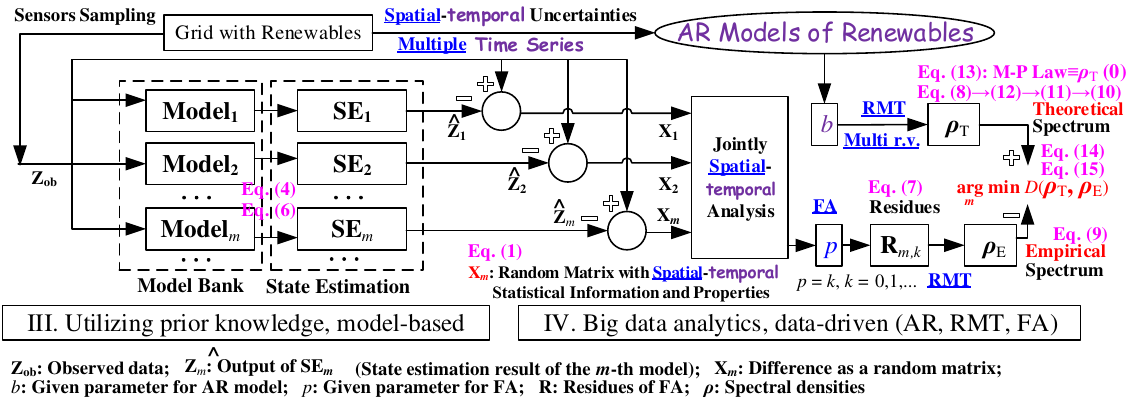}
\caption{Proposed Hybrid Framework}
\label{fig:frame}
\end{figure*}

Fig.~\ref{fig:frame} {summarizes the presented framework} by illustrating how Model Bank, AR, FA, RMT are put together coherently.
The hybrid framework mainly consists of two parts---the model-based part (Sec.~\ref{Sec:ProFor}) and the data-driven part (Sec.~\ref{Sec:Resi}).
The former,  with prior knowledge, converts the observed data into {``difference'' in the form of random matrix}.
Starting from the random matrix and going through \textbf{a rigorous mathematic procedure}, the latter aims to gain insight into the uncertainties through big data analytics, \textbf{with a focus on} the {jointly spatial-temporal analysis and the underlying theories/tools}.

{First}, we build ``bank'' (referring to \cite{singh2010recursive}) to store \textbf{countable} (often a few) virtual models mapping the possible network configurations of a real grid. The bank can be seen as the \textbf{universal set of possible models} among which we try to pick out {the most likely one}. Hence, we need a well-designed metric space---\textbf{{a set together with a metric defined on it}}.

The SE for the models, mainly based on power flow (PF) analysis, is the second step. We make an assumption that each agent on distributed nodes (Agent $i$ on Node $i$ for instance) does collect some local information, such as power usage ($P_{i}$)  and voltage magnitude ($V_{i}$), on its own access point (Node $i$). However, it has \textbf{no prior information} about how it is connected via power lines in the network, {not to mention} power flow on the branch ($P_{i,j}$ and $Q_{i,j}$). The information on $P_{i,j}$ and $Q_{i,j}$ is often a precondition for some SE algorithms \cite{tian2015mixed}, but not for ours. From this aspect, our assumption is \textbf{practical and flexible} for engineering scenarios.

Then we move forwards to the \textbf{difference} $\mathbf X,$ which is modeled as a non-Gaussian random matrix for further big data analytics.
For each bank model (Model $\text M_m$ for instance),  its SE output ($\hat{\mathbf{Z}}_{m}$) does provide a comparison for our observation (${\mathbf{Z}}_{\text{ob}}$), and then the difference $\mathbf X_{m}$ {is defined as}
\begin{equation}
\label{eq:RandomMZ}
\mathbf X_{m}={\mathbf{Z}}_{\text{ob}}-\hat{\mathbf{Z}}_{m}.
\end{equation}

Each $\mathbf X_{m}$ consists of \textbf{multiple time-series}, which can be generally decomposed into four components---the trend, the seasonality, the mutation, and the {randomness}.
Feature extraction of the trend and the seasonality is a well discussed topic in time-series analysis \cite{montgomery2008introduction},
and our previous work \cite{he2015arch} has proposed an RMT-based mutation detection algorithm to handle sudden changes. Here we focus on the randomness.

\subsection{Non-Gaussian Randomness Tools and Related Work}
The randomness component of renewables-derived uncertainties cannot be simply modeled as white noise---\textbf{successive observed data in the form of time-series usually show serial dependence}.
{In order to formally incorporate this (\textit{temporal}) dependent structure}, it is reasonable to explore a general class of models called auto-regressive (AR) models---$x_t\!=\!\sum\nolimits_{i=1}^p{b_ix_{t-1}}\!+\!\epsilon_t$ \cite{bacher2009online}.
From the \textit{spatial} aspect,  FA and RMT are tied together to conduct {jointly temporal-spatial analysis} of the dependence among those multiple time-series.
\begin{enumerate}[1)]
\item Factor Analysis: FA is often used for dimension reduction in high-dimensional datasets \cite{yeo2016random}. Because of the latent constructs (e.g., spatial-temporal independence) lying in the sampling data, FA is preferred to principal component analysis (PCA) \cite{suhr2005principal}. FA has already been successfully applied in various fields such as statistics \cite{fan2011high} and econometrics \cite{dimov2012hidden}. Ref. \cite{pelger2019large} employs FA to handle {high-frequency data} in financial market. In  power system domain, our previous work \cite{shi2019spatio} applies FA to anomaly detection and location with both simulated data and field data.
\item Random Matrix Theory: The entries of a random matrix are random variables and the matrix size is often very large, so RMT is naturally connected with our problem at hand. The goal of RMT is to understand the \textbf{joint eigenvalue distribution}  in the asymptotic regime as the statistic analytics from big data. To our best knowledge, RMT is developed to address this high-dimensional regime since classical statistic theories {apply to low-dimensional regime only}~\cite{qiu2015smart}. Recently, RMT has already been successfully applied in many fields of power system \cite{he2015arch}.
\item ARMA+RMT: This mode is relevant to our big data analytics. Ref. \cite{burda2010random} employs the free random variables (FRV) calculus to calculate the empirical spectral density (ESD) of the sample covariance for several VARMA-type processes.  The derivation is {RMT-based} and {mathematically rigorous}; the {theoretical result} is {nicely matched} against the spectra obtained via Monte Carlo simulations.
\end{enumerate}

\section{Model-based Part Utilizing Prior Knowledge}
\label{Sec:ProFor}
This part aims to convert our observed data into ``difference'' \textbf{in the form of random matrices}.
With PF analysis, the SE output of each bank model is computed as ${\mathbf{Z}}_m$. It supplies a comparison for our observed data ${\mathbf{Z}}_{\text{ob}}$, and hence the difference \textbf{is capable of being defined} (Eq.~\ref{eq:RandomMZ}).
\subsection{Grid Network Operation}
For each node in a power grid, Node $i$ for instance, considering the node-to-ground admittance $y_i$ ($y_i=g_i+\text j \cdot b_i,$ j$=\!\sqrt{-1}$), its active power $P$ and reactive power $Q$ are expressed as:

\begin{normalsize}
\begin{small}
\begin{equation}
\label{eq:PQend}
\begin{aligned}
  & \left\{ \begin{aligned}
  & {{P}_{i}}\!=\!{{V}_{i}}\sum\limits_{k\ne i}{{{V}_{k}}\left( {{G}_{ik}}\text{cos}{{\theta }_{ik}}\!+\!{{B}_{ik}}\text{sin}{{\theta }_{ik}} \right)}\!-\!{{V}_{i}}^{2}\sum\limits_{k\ne i}{{{G}_{ik}}}-{{V}_{i}}^{2}{{g}_{i}} \\
 & {{Q}_{i}}\!=\!{{V}_{i}}\sum\limits_{k\ne i}{{{V}_{k}}\left( {{G}_{ik}}\sin {{\theta }_{ik}}\!-\!{{B}_{ik}}\text{cos}{{\theta }_{ik}} \right)}\!+\!{{V}_{i}}^{2}\sum\limits_{k\ne i}{{{B}_{ik}}}+{{V}_{i}}^{2}{{b}_{i}} \\
\end{aligned} \right. \\
 &      \\
\end{aligned}
\end{equation}
\end{small}
\end{normalsize}

Abstractly, a physical power system obeying Eq.~\eqref{eq:PQend} can be viewed as an analog engine---it takes bus voltage magnitude $V$ and phase angel $\theta$ as \textbf{inputs}, conductance $G$ and susceptance $B$ as \textbf{given parameters}, and ``computes"  active power injection $P$ and reactive power injection $Q$  as \textbf{outputs}.  Thus, the entries of Jacobian matrix $\mathbf J$, i.e. ${\left[ J \right]}_{ij}$, are defined as the partial derivatives of the outputs, $P$ and $Q$, with respect to the inputs, $V$ and $\theta$. All in all, $\mathbf J$ consists of four parts  $\mathbf H, \mathbf N, \mathbf K, \mathbf L$:
\begin{normalsize}
\begin{small}
\begin{equation}
\label{eq:HNKLend}
\left\{ \begin{aligned}
  & {{H}_{ij}}\!={{V}_{i}}{{V}_{j}}\left( {{G}_{ij}}\sin {{\theta }_{ij}}\!-\!{{B}_{ij}}\cos {{\theta }_{ij}} \right)\!-\!{{\delta }_{ij}}\!\cdot\! {{Q}_{i}}\!+\!{{\delta }_{ij}}\!\cdot\! {V}_{i}^2 b_i \\
 & {{N}_{ij}}\!={{V}_{i}}{{V}_{j}}\left( {{G}_{ij}}\cos {{\theta }_{ij}}\!+\!{{B}_{ij}}\sin {{\theta }_{ij}} \right)\!+\!{{\delta }_{ij}}\!\cdot\! {{P}_{i}}\!-\!{{\delta }_{ij}}\!\cdot\! {V}_{i}^2 g_i \\
 & {{K}_{ij}}\!=-{{V}_{i}}{{V}_{j}}\left( {{G}_{ij}}\cos {{\theta }_{ij}}\!+\!{{B}_{ij}}\sin {{\theta }_{ij}} \right)\!+\!{{\delta }_{ij}}\!\cdot\! {{P}_{i}}\!+\!{{\delta }_{ij}}\!\cdot\! {V}_{i}^2 g_i \\
 & {{L}_{ij}}\!={{V}_{i}}{{V}_{j}}\left( {{G}_{ij}}\sin {{\theta }_{ij}}\!-\!{{B}_{ij}}\cos {{\theta }_{ij}} \right)\!+\!{{\delta }_{ij}}\!\cdot\! {{Q}_{i}}\!+\!{{\delta }_{ij}}\!\cdot\! {V}_{i}^2 b_i \\
\end{aligned} \right.
\end{equation}
\end{small}
\end{normalsize}
where $ {{H}_{ij}}\!=\!\frac{\partial {{P}_{i}}}{\partial {{\theta }_{j}}}, {{N}_{ij}}\!=\!\frac{\partial {{P}_{i}}}{\partial {{V}_{j}}}{{V}_{j}}, {{K}_{ij}}\!=\!\frac{\partial {{Q}_{i}}}{\partial {{\theta }_{j}}}, {{L}_{ij}}\!=\!\frac{\partial {{Q}_{i}}}{\partial {{V}_{j}}}{{V}_{j}}$.

\subsection{Power Flow Analysis}
PF analysis deals mainly with the calculation of steady-state system status, i.e., voltage magnitude $V$ and phase angel $\theta$, on each network bus, for a given set of variables such as load demands, under certain assumptions such as in a balanced system operation \cite{gomez2018electric}.
Conventional PF analysis is model- and assumption-based. That is to say, the information of network topology $\mathbf Y$ is a \textbf{prerequisite} for the calculation, and the input (output) variables need to be \textbf{preset} as one of the following three categories:
\begin{itemize}
\item $P$ and $V$ ($Q$ and $\theta$) for voltage controlled bus/$PV$ bus;
\item $P$ and $Q$ ($V$ and $\theta$) for load bus/$PQ$ bus;
\item $V$ and $\theta$ ($P$ and $Q$) for reference bus/slack bus.
\end{itemize}

Consider a power system with $n$ buses, among which there are  $m$ $PV$ buses,  $l$ $PQ$ buses, and $1$ slack bus ($n\!=\!l\!+\!m\!+\!1$). Starting with Eq. \eqref{eq:PQend}, PF functions is formulated as Eq.~\eqref{eq:J1}.

\begin{equation}
\label{eq:J1}
\mathbf{y}\!:=\! \left[ \begin{matrix}
   {{P}_{1}}  \\
   \vdots   \\
   {{P}_{n-1}}  \\
   {{Q}_{m+1}}  \\
   \vdots   \\
   {{Q}_{n-1}}  \\
\end{matrix} \right]\!=\!\bm f\left[ \begin{matrix}
   {{\theta }_{1}}  \\
   \vdots   \\
   {{\theta }_{n-1}}  \\
   {{V}_{m+1}}  \\
   \vdots   \\
   {{V}_{n-1}}  \\
\end{matrix} \right]\!=:\!\bm f\left( \mathbf{x} \right)
\quad \mathbf{J}\!=\!\left[ \begin{matrix}
   \frac{\partial {{y}_{1}}}{\partial {{x}_{1}}} & \cdots  & \frac{\partial {{y}_{1}}}{\partial {{x}_{K}}}  \\
   \vdots  & \ddots  & \vdots   \\
   \frac{\partial {{y}_{K}}}{\partial {{x}_{1}}} & \cdots  & \frac{\partial {{y}_{K}}}{\partial {{x}_{K}}}  \\
\end{matrix} \right]
\end{equation}
where $:=$ is the assignment symbol in computer science.

Eq.~\eqref{eq:J1} builds a differentiable mapping function $\bm f\!:\!\mathbf{x}\!\in\! {{\mathbb{R}}^{K}}\!\to\! \mathbf{y}\!\in\! {{\mathbb{R}}^{K}}$. It consists of $K\!=\!2n\!-\!m\!-\!2$ equations,  from the same number ($m\!+\!2l\!=\!K$) state variables, $\theta$ and $V$, to the power injections, $P$ and $Q$.
Following Eq. \eqref{eq:HNKLend},  $\mathbf J$ is calculated as a $K\!\times \!K$ matrix:
\begin{equation}
\label{eq:defJ}
\mathbf{J}= \left[ \begin{array}{*{35}{l}}
   {{\left[ \mathbf H \right]}_{n-1, n-1}} & {{\left[ \mathbf N \right]}_{n-1, n-m-1}}  \\
   {{\left[ \mathbf K \right]}_{n-m-1, n-1}} & {{\left[ \mathbf L \right]}_{n-m-1, n-m-1}}  \\
\end{array} \right]
\end{equation}

To formulate the \textbf{linear approximation} process that the system operation point shifts from $(\mathbf{x}^{(k)},\mathbf{y}^{(k)})$ to $(\mathbf{x}^{(k\!+\!1)},\mathbf{y}^{(k\!+\!1)})$, the iteration is set as follows:
\begin{equation}
\label{eq:XJ}
{{\mathbf{x}}^{\left( k+1 \right)}}:={{\mathbf{x}}^{\left( k \right)}}+{\mathbf {J}}^{-1}\left( {{\mathbf{x}}^{\left( k \right)}} \right)\left( {{\mathbf{y}}^{\left( k+1 \right)}}-{{\mathbf{y}}^{\left( k \right)}} \right)
\end{equation}

The iteration depicts how to update the state variables from ${\mathbf{x}}^{\left( k \right)}$ to ${\mathbf{x}}^{\left( k+1 \right)}$.
${\mathbf{y}}^{\left( k \right)}$ and ${\mathbf{x}}^{\left( k \right)}$ are known quantities under our assumption in Sec.~\ref{Sec:Assumpt}.
${\mathbf{y}}^{\left( k+1 \right)}$, according to Eq.~\eqref{eq:J1}, is the desired $P, Q$ on $PQ$ buses and desired $P$ on $PV$ buses\footnote{For $PQ$ buses, neither $V$ nor $\theta$ are fixed; they are state variables that need to be estimated.  For $PV$ buses, $V$ is fixed, and $\theta$ needs to be estimated.}. ${\mathbf{x}}^{\left( k+1 \right)}$ is the state variables that need to be estimated through the iteration in this expression (Eq. \ref{eq:XJ}).

The above model-based deterministic PF analysis  \textbf{is not always reliable in practice}, since the {network topology} $\mathbf Y$, and the operation points $(\mathbf{x}^{(k)},\mathbf{y}^{(k)})$ are required \textbf{to be of high precision and up-to-date}. These requirements, unfortunately, are often \textbf{unrealistic} as mentioned in Sec \ref{Introd}.

\subsection{Model Bank}
\label{Sec:ModelBankk}

\begin{figure}[htbp]
\centering
\includegraphics[width=0.46\textwidth]{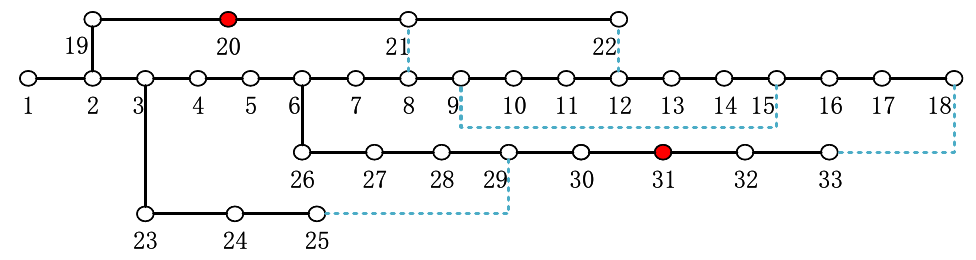}
\caption{IEEE 33-bus Network}
\label{fig:ieee33p1}
\end{figure}

During the daily operation of a distribution grid, its topology may be partially reconfigured due to maintenance or emergency/optimal operation. Taking IEEE 33-bus network for instance, the network topology is shown in Fig.~\ref{fig:ieee33p1}. It is a 12.66-kV distribution grid system including a substation and 37 branches.
The normally closed branches are represented by solid lines, and normally opened ones by dashed lines.

With the pair switch of these normally closed/opened branches, the grid has `countable' possible network configurations.
In practice, however, it is reasonable to study \textbf{only a few models} even for a large system, since the network configuration of a particular grid must be confined to several typical models.
We employ the concept of `bank' referring to \cite{singh2010recursive} to store them, and the deterministic PF analysis works out the SE results of these models as $\hat{\mathbf{Z}}_{m}$. These SE outputs $\hat{\mathbf{Z}}_{m}$ allow of the comparison with our observed data ${\mathbf{Z}}_{\text{ob}}$, and hence the difference $\mathbf X$ {is capable of being defined} as a random matrix (Eq.~\ref{eq:RandomMZ}).
In this way, the TI task is converted into \textbf{a matching problem under certain metric space}. The design of the metric space will be discussed in Sec.~\ref{Sec:MetricDesign}.

\section{High-Dimensional  Analysis with RMT and FA}
\label{Sec:Resi}

Our motivation arises from the fact that  \textbf{the renewables-derived uncertainties cannot be simply modeled as white noise}. It does contain much (latent) structural information, especially when there is an extra bias caused by a certain (although maybe unknown) poor assumption or negligence. For this purpose, FA is employed in our framework.
The entries of the resultant matrix are \textbf{random variables and large in size}, so RMT is naturally relevant to the problem \cite{he2015arch}.

\subsection{RMT-based Problem Formulation}
The goal of RMT is to understand \textbf{joint eigenvalue distribution} in the asymptotic regime as big data analytics.
The spectrum of a covariance matrix generally consists of two parts: A few spikes/outliers and the bulk. The former represents \textbf{common factors} that mainly drive the features, and the latter represents \textbf{unique factors or error variation} that arise from \textbf{idiosyncratic noise}.
For the noise part, we consider a \textbf{minimum distance} between two spectral densities---a \textbf{\textit{theoretical}} one $\bm \rho_{\text T}$ from an ideal structure model, and an \textbf{empirical} one $\bm \rho_{\text E}$ relevant to the (multiple time-series) observed data.

\subsection{Factor Analysis Formula}
In dealing with high-dimensional datasets, FA is often used for dimension reduction in sampling data with underlying constructs that cannot be measured directly \cite{yeo2016random, suhr2005principal}.

Regarding empirical data  $\mathbf{X}\in {{\mathbb{R}}^{N\times T}}$, FA is formulated as
\begin{equation}
\label{eq:FactM}
\mathbf{X}={{\mathbf{L}}^{\left( p \right)}}{{\mathbf{F}}^{\left( p \right)}}+\mathbf{R}.
\end{equation}
where  $\mathbf{F}\in {{\mathbb{R}}^{p\times T}}$ is a matrix of common factors, $\mathbf{L}\in {{\mathbb{R}}^{N\times p}}$  is factor loadings,  $p$ is factor numbers, and $\mathbf{R}\in {{\mathbb{R}}^{N\times T}}$  is residues, also called unique factors or error variation.

Eq.~\eqref{eq:FactM} enables to decompose observed data $\mathbf X$ into \textbf{systematic information} and \textbf{idiosyncratic noise}. Usually, only $\mathbf X$ is \textbf{observable}, $\mathbf L$ is composed of the first $p$ principal components of $\mathbf{X}$, $\mathbf{F}\!=\!{{\left( {{\mathbf{L}}^{\text{T}}}\mathbf{L}  \right)}^{-1}}{{\mathbf{L}}^{\text{T}}}\mathbf{X}$, and $\mathbf{R}\!=\!\mathbf{X}-{{\mathbf{L}}}{{\mathbf{F}}}.$

We focus on residues $\mathbf R$,  which may contain some latent constructs and statistical information.
Instead of regarding $\mathbf R$ as Gaussian noise a priori, we assume that there are \textbf{cross- and auto-correlated structures}. Without loss of generality, $\hat{\mathbf R}$ is represented as
\begin{equation}
\label{eq:crossautocorrelated}
\hat{\mathbf R} \!=\! {\mathbf A_{N}^{1/2}}\mathbf{\epsilon}{\mathbf B_{T}^{1/2}}	
\end{equation}
where $\mathbf{\epsilon}$ is an $N\!\times\! T$ Gaussian matrix with independent and identically distributed (i.i.d.) random entries, $\mathbf A_N$ and $\mathbf B_T$ are $N\!\times\! N$ and $T\!\times\! T$ symmetric non-negative definite matrices, representing cross- and auto- covariances, respectively. %\footnote{This is not the most general model, since cross- and auto-covariance contributions are decoupled: $\text{cov}(R_{it},R_{js}) \!=\!{A_N}_{ij}{B_T}_{ts}$.}.
Eq.~\eqref{eq:crossautocorrelated} leads to \textbf{a separable sample covariance matrix} in the sense that ${\bf A}_N$ and ${\bf B}_T$ are separable. This structural assumption of separability is \textbf{a popular assumption} in the analysis of spatial-temporal data~\cite{ding2019spiked}.
%Although this assumption does not allow for spatial-temporal interactions in the covariance matrix, in many real data applications, the covariance matrix \textbf{can be well approximated} using separable covariance matrices by solving a nearest Kronecker product for a space-time covariance matrix problem.
Although this assumption does not allow for spatial-temporal interactions in the covariance matrix, in many real data applications, the covariance matrix \textbf{can be well approximated} using separable covariance matrices for a space-time covariance matrix problem.

%In our task, we restrict the matrix structures of ${\mathbf A_{N}}$ and ${\mathbf B_T},$ so that they are completely defined by simple parameter sets, i.e., $\bm \theta \!=\! (\bm \theta_{\mathbf A_N}, \bm \theta_{\mathbf B_T})$ that are to be estimated along with factor numbers $p$.
%For example, a simple case is that each residue has the same cross-correlation\footnote{We assume each time-series, $R_{it} (t\!=\!1,\cdots,T)$, is normalized and has a zero Mean and an unit Variance.}, called $\beta$, to other residues, and each residue has an exponentially decaying temporal auto-correlations with a parameter $\tau$. Then two parameters $\bm \theta_{\mathbf A_N}\! =\!\beta$ and $\bm \theta_{\mathbf B_T}\! =\!\tau,$ completely determine $\mathbf A_N$ and $\mathbf B_T$, since $\mathbf A_N\!=\!\{{A}_{ii}\!=\!1,\, {A}_{ij,i \!\ne\! j}\!=\!\beta, \; i,j\!=\!1,\cdots,N\}$ and $\mathbf B_T\!=\!\{{B}_{st}\!=\!e^{-|s-t|/\tau}, \; s,t\!=\!1,\cdots,T\} \  (e\!\approx \!2.718).$

\subsection{FA Estimation Based on Spectrum Analysis}
\label{sec:fmsa}
Now the objective of the mentioned matching problem is to match the spectral density $\bm \rho_{ \text{E}}$ against $\bm \rho_{ \text{T}}.$

The former  $\bm \rho_{ \text{E}}$ means the \textbf{ESD} of the covariance matrix of residues $\mathbf R$ constructed from \textbf{empirical data}. It can be controlled by the $p$ number of common factors to be removed following Eq.~\eqref{eq:FactM}. It is defined as \cite{rogers2010new}
\begin{equation}
\label{eq:ESDdef}
\bm\rho_{ \text{E}}(\lambda)=\frac{1}{N} \sum_{i=1}^{N} \delta\left(\lambda-\lambda_{i}^{(\mathbf C_N)}\right)
\end{equation}
where $\{\lambda_{i}^{(\mathbf C_N)}\}_{i=1}^N$ is the eigenvalues of  $\mathbf C_N\!=\! \frac{1}{T}\mathbf R \mathbf R^{\text T}$, and $\delta$ is the Dirac delta function.

The latter $\bm \rho_{ \text{T}}$ means the \textbf{theoretical spectral density} of the ideal covariance matrix $\hat{\mathbf C}_N$ with the assumed structural model, i.e.,
$\hat{\mathbf C}_N\!=\! \frac{1}{T}\hat{\mathbf R} \hat{\mathbf R}^{\text T} \!=\! \frac{1}{T}{\mathbf A_{N}^{1/2}}\mathbf {\epsilon}{\mathbf B_T}{\mathbf{\epsilon}^{\text T}}{\mathbf A_{N}^{1/2}}$ (Eq.~\ref{eq:crossautocorrelated}).
Assuming a parsimonious matrix structure of $\mathbf A_N$ and $\mathbf B_T,$ which is determined by only a small parameter set $\bm \theta.$
Mathematically motivated by the result of \cite{zhang2006spectral},  the spectral density of $\hat{\mathbf C}_N$, under certain assumptions, converges to a certain \textbf{limiting distribution} $\bm \rho_{\text{T}}(\bm \theta)$, as the size $N$ tends to infinity.

\subsection{Simplified Model on Covariance Structures of Residues}
\label{sec:crosscorr}
A difficulty lies in the calculation of the limiting density,  $\bm \rho _{\text{T}}(\bm \theta)$, for general $\bm \theta \!=\! (\bm \theta_{\mathbf A_N}, \bm \theta_{\mathbf B_T})$.
The actual calculation of $\bm \rho _{\text{T}}(\bm \theta)$ is quite complex, which makes the implementation difficult. A recent study of \cite{burda2010random}, fortunately, provides the direct derivation of this limiting spectral density using free random variable (FRV) techniques. They particularly present \textbf{analytic forms} when the time-series follow ARMA processes. In our task, we employ these techniques to calculate $\bm \rho _{\text{T}}(\bm \theta).$  First, two assumptions are made:
\begin{enumerate}[I.]
\item The cross-correlations of $\hat{\mathbf R}$ are effectively eliminated by removing $p$ factors, and therefore $\hat{\mathbf R}$ has sufficiently negligible cross-correlation: $\mathbf A_N \!\approx\! \mathbf I_{N\times N}.$
\item The auto-correlations of $\hat{\mathbf R}$ are exponentially decreasing, i.e., $\{ B_T \}_{ij} = b^{|i-j|}$, with $|b|<1.$\footnote{This is equivalent to modeling residues as
an $\text{AR}(1)$ process: $\hat{R}_{it} \!=\! b\hat{R}_{i,t-1} \!+\! \xi_{it}$, where $\xi{}\! \sim \!\mathcal{N}\left( 0,1-{{b}^{2}} \right) $ so that the variance of $\bm \hat{R}_t$ is 1.}
\end{enumerate}

Under the two assumptions, we can conduct spectrum analysis of the simplified model, and thus $\bm \rho _{\text{T}}(b)$ is capable of being computed.  The major steps are briefly given as follows:
\begin{enumerate}[1.]
\item The mean spectral density  can be derived from the Green's function $G(z)$ by using the Sokhotsky's formula:
\begin{equation}
\label{Eq:inverse_green_function}
\begin{aligned}
  \bm \rho_{\text {T}}(\lambda) = -\frac{1}{\pi} \lim\limits_{\varepsilon\rightarrow 0^{+}} \text{Im} G(\lambda + i\varepsilon)
\end{aligned}.
\end{equation}
\item  The Green's function $G(z)$ can be obtained from the moments' generating function $M(z):$
\begin{equation}
\label{Eq:MzGz}
\begin{aligned}
  G(z) =  \frac {M(z)+1}{z}, \qquad |z| \ne 0
\end{aligned}.
\end{equation}
\item $M(z)$ can be found by solving the polynomial equation:
\begin{equation}
\label{Eq:polynomial}
\begin{aligned}
  a^4c^2M^4+2a^2c(-(1+b^2)z+a^2c)M^3+((1-b^2)^2z^2 \\
  -2a^2c(1+b^2)z+(c^2-1)a^4)M^2-2a^4M-a^4=0
\end{aligned},
\end{equation}
where $a = \sqrt{1-b^2}$, and $c = \frac{N}{T}.$
\end{enumerate}

It is worth mentioning that when $b\!=\!0$, Assumptions \uppercase\expandafter{\romannumeral 1}  \& \uppercase\expandafter{\romannumeral 2} imply that $\hat{\mathbf R}$ is a standard Gaussian matrix with i.i.d. random elements, and its spectral density is marked as $\bm \rho _{\text{T}}(0)$.
On the other side, Marchenko-Pastur Law says that for a  Laguerre unitary ensemble (LUE) matrix $\mathbf{\Gamma}\!\in\! \mathbb C^{N\!\times\!T} \left( {c}\!=\!{N/T}\le 1\right)$, its  spectral density  ${\bm g_{\text{MP}}}\left( x \right)$ does follow M-P Law \cite{marvcenko1967distribution}:
\begin{equation}
\label{eq:D2}
{\bm g_{\text{MP}}}\left( x \right) = \frac{1}{{2\pi cx}}\sqrt {\left( {x - s_1} \right)\left( {s_2 - x} \right)}, \qquad x \in \left[ { s_1,s_2} \right]
\end{equation}
where $s_1 = {\left( {1 - \sqrt c  } \right)^2} \text{ and } s_2 = {\left( {1 + \sqrt c  } \right)^2}$.

\begin{figure}[htbp]
\centering
\subfloat[${c}\!=\!{N/T}\!=\!0.6$]{\label{fig:MPc06}
\includegraphics[width=0.20\textwidth]{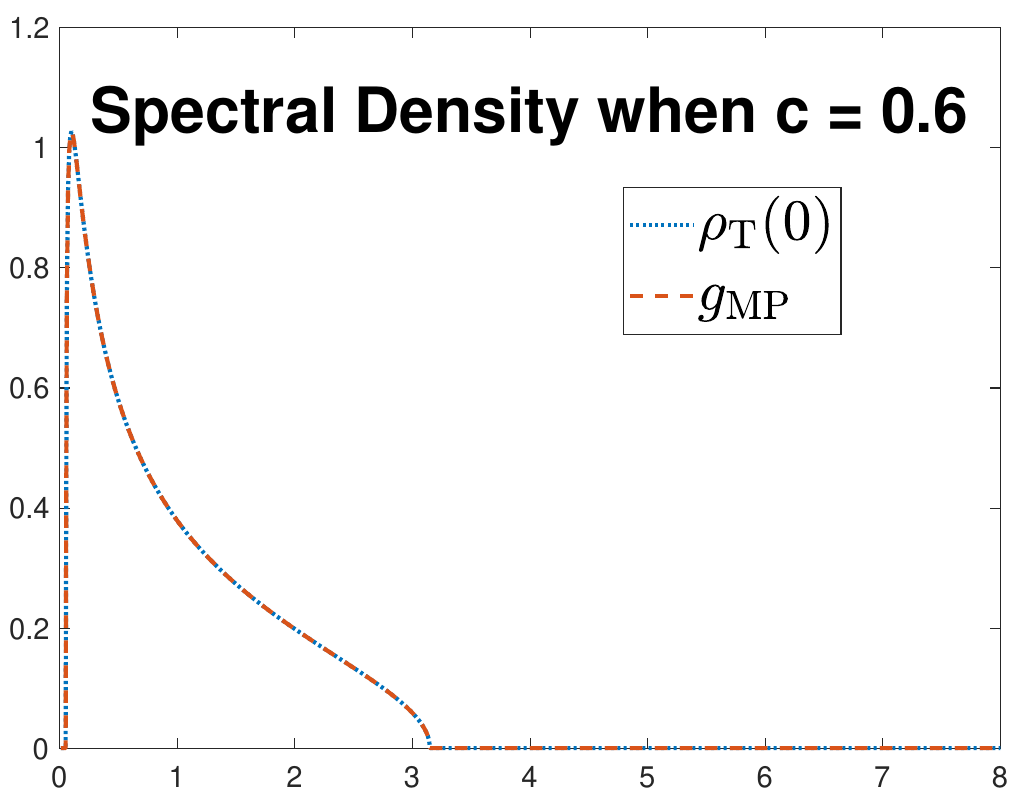}}
\subfloat[${c}\!=\!{N/T}\!=\!0.3$]{\label{fig:MPc03}
\includegraphics[width=0.28\textwidth]{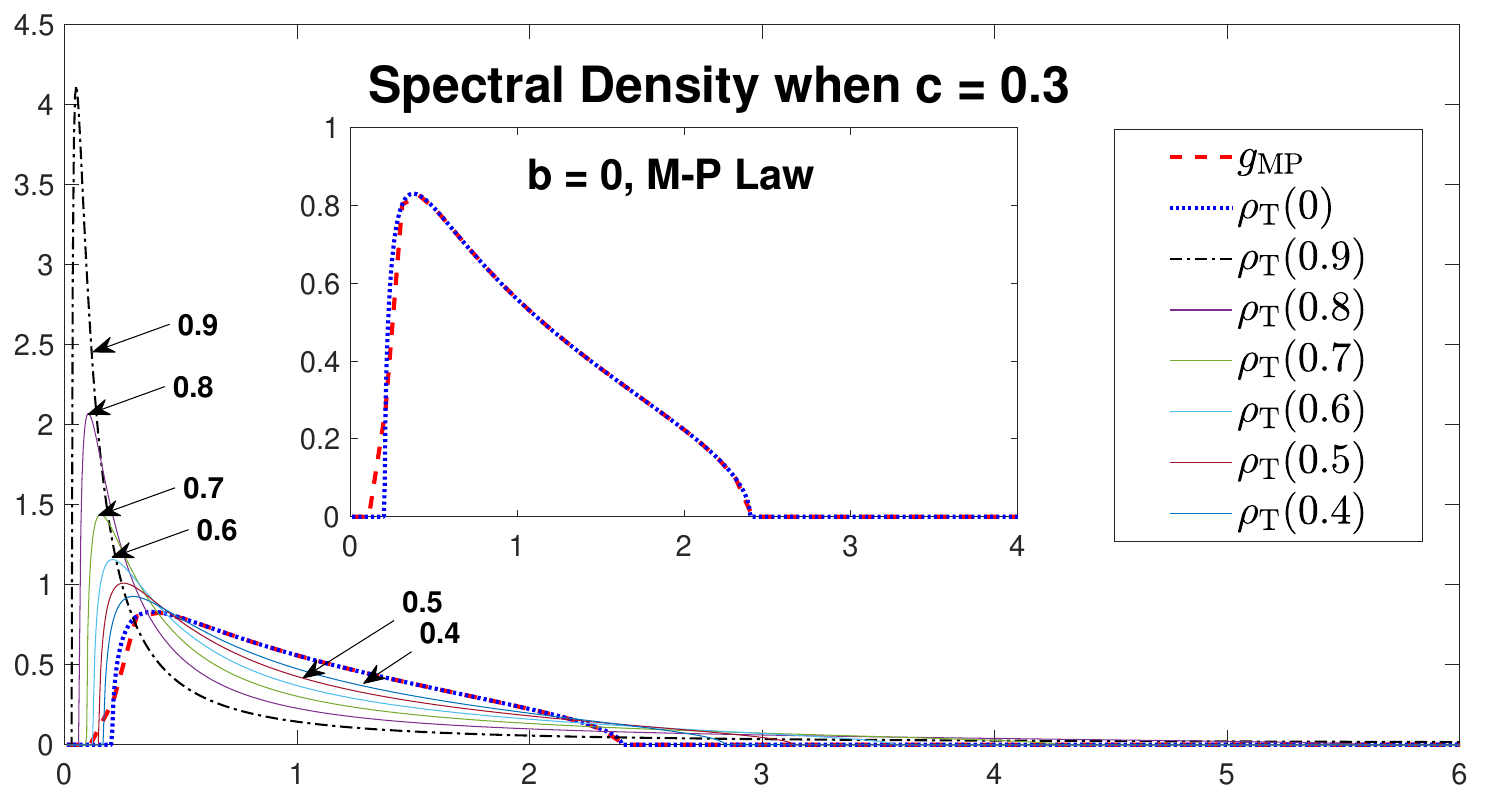}}
\caption{Spectral Density of $\bm \rho _{\text{T}}(b)$ and $\bm g_{\text{MP}}$}
\label{fig:ARMP}
\end{figure}

The two spectral densities should be equivalent, i.e. $\bm \rho _{\text{T}}(0)$  is equivalent to $\bm g_{\text{MP}}$. Fig.~\ref{fig:ARMP} displays this phenomenon.

Fig.~\ref{fig:MPc03} also tells that the theoretical spectral densities $\bm \rho _{\text{T}}(b)$ are \textbf{distinguishable} with different coefficients $b$ in the AR model.
This property implies that the  (latent) coefficients $b$ offers \textbf{good potential for the metric space construction}.
With the help of metric space, the randomness component of the observed data is able to be addressed {from the view of spectrum analysis}.

\subsection{Metric Space Designing}
\label{Sec:MetricDesign}

To design a metric space for solving the mentioned match problem, we need to assign a set and then define a distance function (metric) on it.
What we have in practice are the observed data $\mathbf{Z}_\text{ob}$ in the form of multiple time-series, and SE outputs $\hat{\mathbf{Z}}_m$ derived from $\mathbf{Z}_{\text{ob}}$ and Model $\text M_m$.
The difference (Eq.~\ref{eq:RandomMZ}: $\mathbf X_{m}={\mathbf{Z}}_{\text{ob}}-\hat{\mathbf{Z}}_{m}$) is the first and most obvious choice for us to extract some statistical information from.

Before designing the metric space, let us look through those conventional statistics indexes,  e.g., first/second moment (mean/variance).
We have already argued that the profile of renewables-derived uncertainties does follow AR models.
Mean and variance contain enough statistical information for an i.i.d. Gaussian random variable, but insufficient for an AR model, not to mention multiple AR processes (temporal aspect) on  those connected distributed access points (spatial aspect).

Some more powerful tools are needed to map the \textbf{difference} $\mathbf{X}_m$, which consists of a large number of random variables, into some indicator within a well designed metric space.
The proposed hybrid framework (Fig.~\ref{fig:frame}) conducts jointly temporal-spatial analysis of $\mathbf{X}_m$ as follows:
First, $\mathbf{X}$ is converted into $\mathbf{R}$  with a given $p$ (Eq.~\ref{eq:FactM}), and then the ESD $\bm\rho_{ \text{E}}$ is calculated (Eq.~\ref{eq:ESDdef}).
On the other hand,  with a given coefficient~$b$, the theoretical spectral density $\bm\rho_{ \text{T}}(b)$ is capable of being computed (Eq.~\ref{eq:crossautocorrelated}$\rightarrow$\ref{Eq:polynomial}$\rightarrow$\ref{Eq:MzGz}$\rightarrow$\ref{Eq:inverse_green_function}). For convenience, \textbf{the metric distance such as Jensen-Shannon divergence} can be studied:
\begin{equation}
\label{eq:Matric}
 d(\mathbf{Z}_\text{ob}, \hat{\mathbf{Z}}_{m}) \! =\! |\mathbf{X}_m|_{\mathcal D}\! =\! \mathcal D(\bm{\rho}_{\text{T}}(b), \bm\rho_{ \text{E}}(p))\! =\! \!\sum\nolimits_i p_i \mathcal D_{\text{JS}}(a_i, b_i)
\end{equation}
where $\mathcal D_{\text{JS}}(a, b)\! =\! a\log a\!+\!b\log b \!-\! 2v\log v$ with  $v\!=\!\frac{a+b}{2}$.

With the metric space design, the TI task is converted into a convex optimization problem
\begin{equation}
\label{eq:tiOptim}
\text{arg}\min\limits_{m} d(\mathbf{Z}_\text{ob}, \hat{\mathbf{Z}}_{m}) \!=\! \text{arg}\min\limits_{m} \! \mathcal D(\bm{\rho}_{\text{T}}(b), \bm\rho_{ \text{E}}(p)).
\end{equation}
The convex optimization can be readily calculated using modern software toolbox such as CVX.

\section{Case Studies}
\label{Sec:Cases}

\subsection{Case Background and Model Bank}
IEEE 33-bus Network (Fig.~\ref{fig:ieee33p1}) is used to validate our proposed hybrid framework.
Considering a sampling dataset with 1440 observations (4 hours with a 0.1 Hz sampling rate).
This observation leads to the empirical dataset $\mathbf{Z}_\text{ob}$, which consists of local sample data from 33 access points.
Following Sec.~\ref{Sec:Assumpt}, it is assumed that there is no prior information about the power flow on the connected branch ($P_{i,j}$ and $Q_{i,j}$).

\begin{figure}[htbp]
\centering
\subfloat[$\mathbf{P}_\text{real}$: Load Behavior  in IEEE 33-bus Network]{\label{fig:Power}
\includegraphics[width=0.46\textwidth]{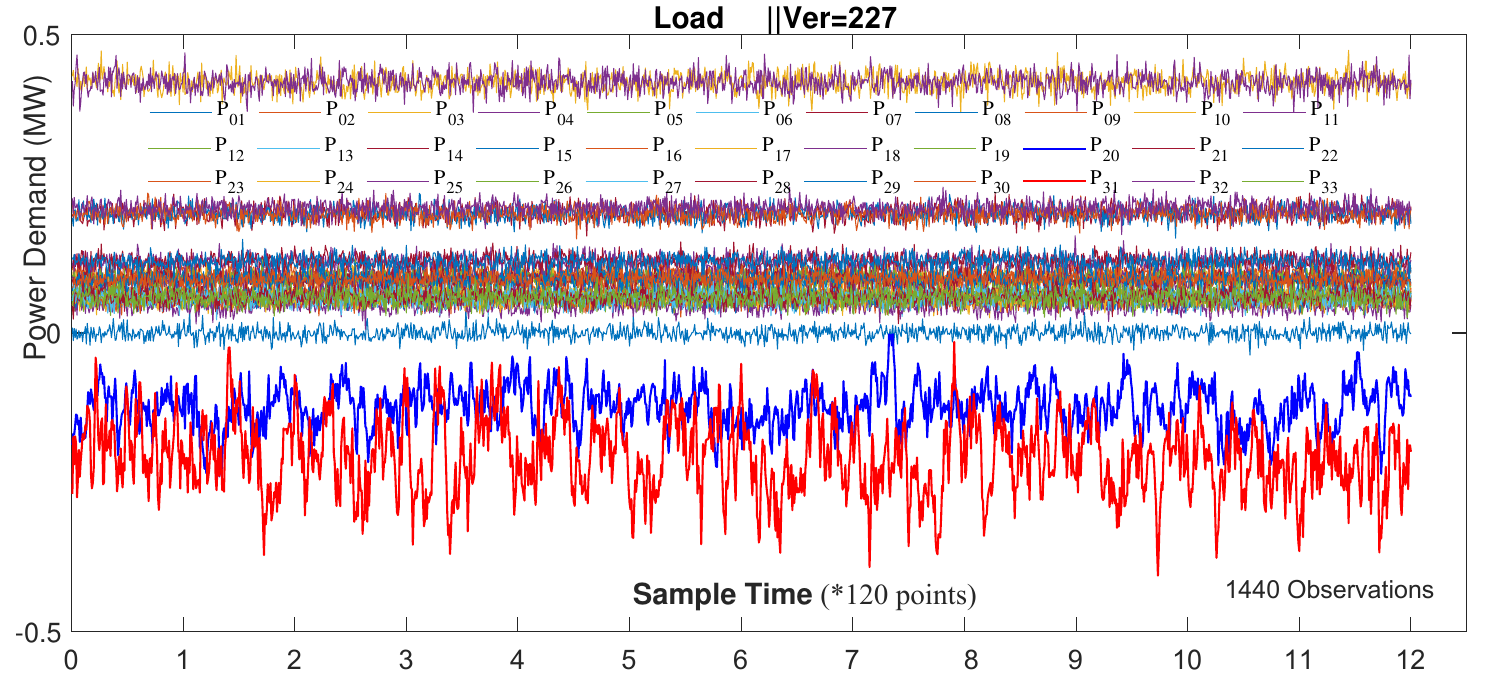}}

\subfloat[$\hat{\mathbf{V}}_{1}$: Voltage Magnitude of Model 1]{\label{fig:Voltage}
\includegraphics[width=0.46\textwidth]{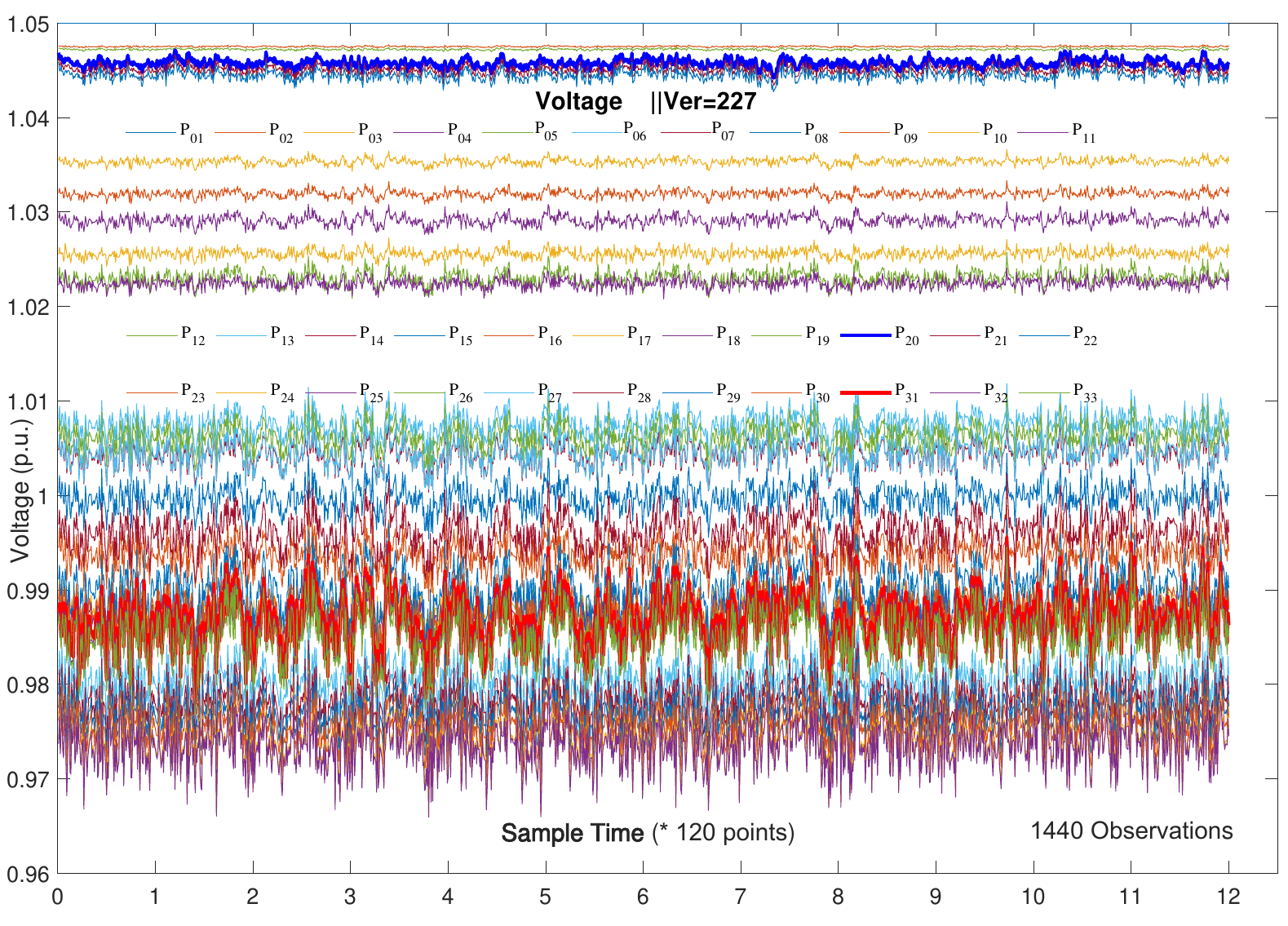}}
\caption{Dataset from 33 Points and 1440 Observations}
\label{fig:Zreal}
\end{figure}

Fig \ref{fig:Power} depicts the active power generation/consumption at each node (${P_i\in\mathbf{P}}_\text{ob}\!\subset\!{\mathbf{Z}}_{\text{ob}}$).
For Node 20 and Node 31, the curves are of {high variation} derived from the behavior of some \textbf{wind speed data in practice}.
For other nodes, however, the curves are stationary since the profile of routine power usages is relatively smooth.
It is noteworthy that we only discuss the randomness component as mentioned in Sec. \ref{Sec:Framework}.
%AR model and RMT are employed, respectively, to handle the autocorrelation and  other randomness feature \textbf{in high-dimensional space}.

\begin{figure}[htbp]
\centering
\includegraphics[width=0.46\textwidth]{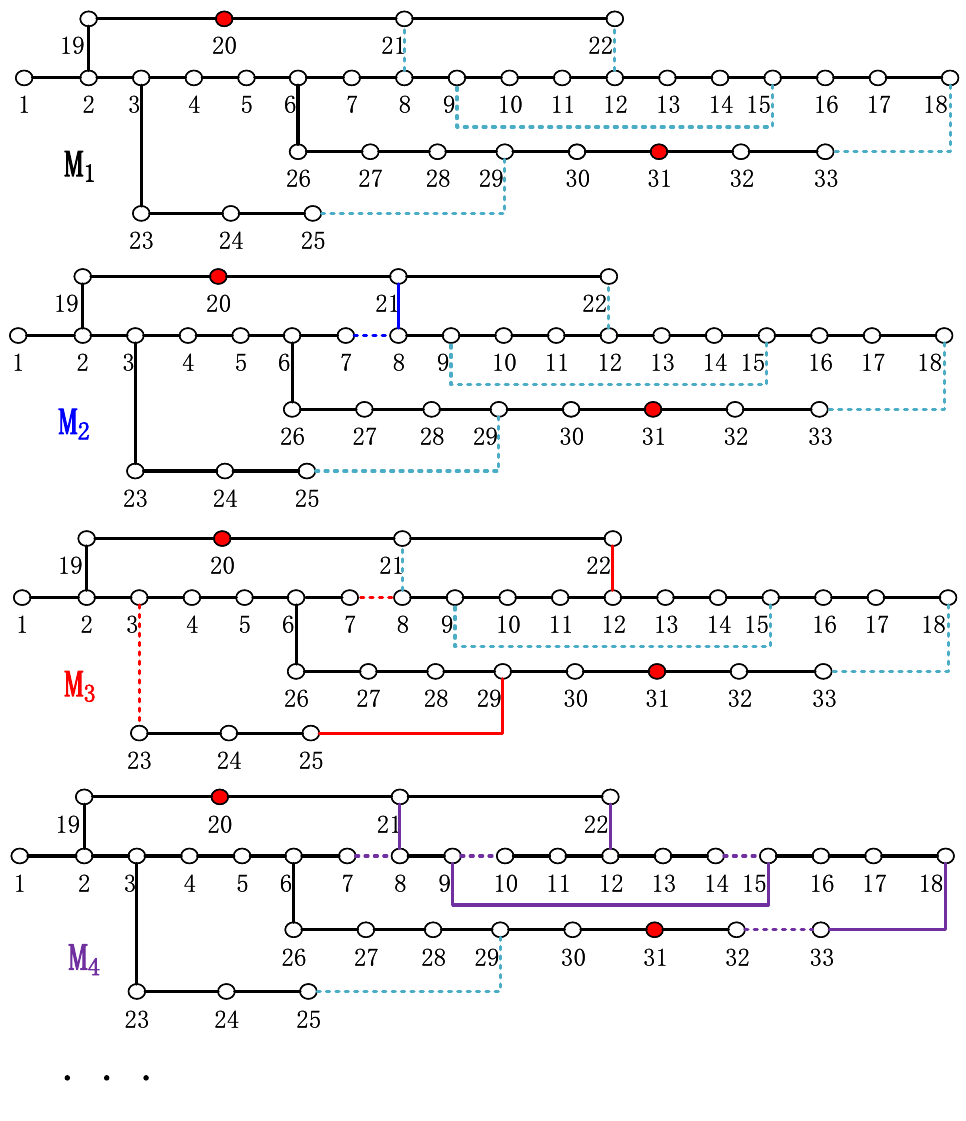}
\caption{Models Stored in the Model Bank}
\label{fig:modebank}
\end{figure}

The physical grid only has numerous possible operation models (Sec. \ref{Sec:ModelBankk}), and we arrange them to form the model bank (Fig \ref{fig:modebank}).
Through parallel PF analysis, we test each model, e.g. Model $\text M_m$, and work out its SE result $\hat{\mathbf{Z}}_{m}$.
Fig \ref{fig:Voltage} depicts the voltage magnitudes of Model $\text M_1$($\hat{\mathbf{V}}_1\!\subset\!\hat{\mathbf{Z}}_{1}$).

%The vibration of $\hat{\mathbf V}_1$ on each node, e.g. Node $k$, does follow AR model as aforementioned.
The low-dimensional statistics Mean $\mu$  and Variation $\sigma$ contain enough statistical information about Gaussian variables, but not about the renewables-derived randomness $\hat{\mathbf{V}}_1$, of which multiple AR time-series contribute a major part.
Moreover, Mean $\mu$ is \textbf{vulnerable to fixed measurement error}.
To address those renewables-derived uncertainties is the primary motivation for our proposed framework.
%According to our framework (Fig.~\ref{fig:frame}), we conduct jointly temporal-spatial analysis to these multiple time-series, and try to gain some big data analytics according to Sec. \ref{Sec:Resi}.

\subsection{Case Designing}
\label{Sec:CaseDsigning}

We assume that at time point $t\! =\! 720$, due to some reason there is an operation model transformation from Model $\text M_1$ to $\text M_2$---the  system operates under $\text M_1$ during $ 0 \!\sim\! 720$, and $\text M_2$ during $721 \!\sim\!1440$.
We also take the {measurement error} into account, and regard it as a Gaussian random variable $E$, whose statistical properties can be fully described by mean $\mu_E$ and standard deviation $\sigma_E$.

Our previous work \cite{he2016les} has already shown that the \textbf{fixed measurement error $\mu_E$ has no influence} to the RMT-based analysis and indicator at all. Therefore we only need to consider $\sigma_E$. Referring to  \cite{ma2012development}, it is supposed that $\sigma_E\!=\!0.005$ p.u.---the standard deviation of the measurement errors is 0.5\%.
The uncertainties caused by renewables and measurement errors together may {significantly influence} the statistical properties of observed data, thereby disabling TI performance.

\begin{figure}[htbp]
\centering
\includegraphics[width=0.48\textwidth]{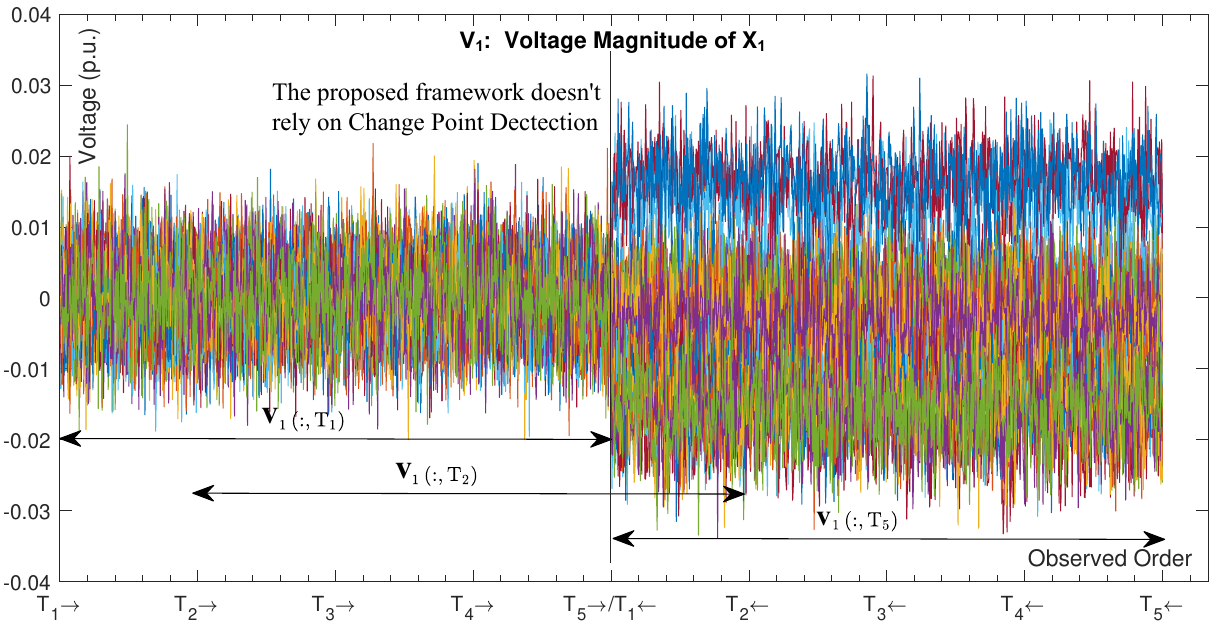}
\caption{$\mathbf{V}_1$: Voltage Magnitude Component of Difference $\mathbf{X}_1$}
\label{fig:ResiduesX1}
\end{figure}

When both $\mathbf{Z}_\text{ob}$, the observed data,  and  $\hat{\mathbf{Z}}_{m}$, the SE output of Model $\text M_m$, are known a priori, so is their difference $\mathbf{X}_m$.  As the reasons given in our previous work \cite{he2015arch}, only voltage magnitude $\mathbf V_m\!\subset\!\mathbf X_m$ is discussed.
Furthermore, we keep each observation duration 720 sampling points and thus divide the whole observation into 5 periods: $\text{T}_1$ ($ 1 \!\sim\! 720$), $\text {T}_2$ ($ 181 \!\sim\! 900$), $\text {T}_3$ ($ 361 \!\sim\! 1080$), $\text {T}_4$ ($ 541 \!\sim\! 1260$),  and $\text {T}_5$ ($ 721 \!\sim\! 1440$).
We use $\mathbf V_m(:, \text{T}_j)$ to represent the voltage difference on all the 33 nodes during $\text {T}_j$, which can be denoted as $\mathbf{V}_{m\_j}$ when there is no ambiguity. Fig.~\ref{fig:ResiduesX1} shows the voltage magnitude difference in each period for Model $\text M_1$:  $\mathbf{V}_{1\_1}$, $\mathbf{V}_{1\_2}$, $\cdots$, $\mathbf{V}_{1\_5}$.

\subsection{Regression-based TI and its Failure when Uncertainties are not Well Addressed}
\label{sec:RegandF}

We test the TI performance by employing Jacobian matrix $\mathbf J$ (Eq.~\ref{eq:defJ}), a matrix variate which is strongly associated with network topology $\mathbf Y$.
From Eq.~\eqref{eq:J1}, the estimation of $\mathbf{J}$  can be naturally formulated as \textbf{a regression problem}.
Under {fairly general conditions},  the target $\mathbf {J}$, according to Eq.~\eqref{eq:HNKLend},  {keeps nearly constant within some duration}, called $\Delta t$, due to the stability of the system, or concretely, of variables $V, \theta, Y$. During  $\Delta t$, considering $T$ times observation at time instants $t_i$, $(i\!=\!1,2,\cdots,T, t_T\!-\!t_1\!=\!\Delta t)$, we acquire the operation data points in the form of $(\mathbf{x}^{(i)},\mathbf{y}^{(i)})$.

\begin{figure}[ht]
\centering
\subfloat[Mean:  $\mathbf J_{\text{Mean}}$]{\label{fig:Jmean}
\includegraphics[width=0.24\textwidth]{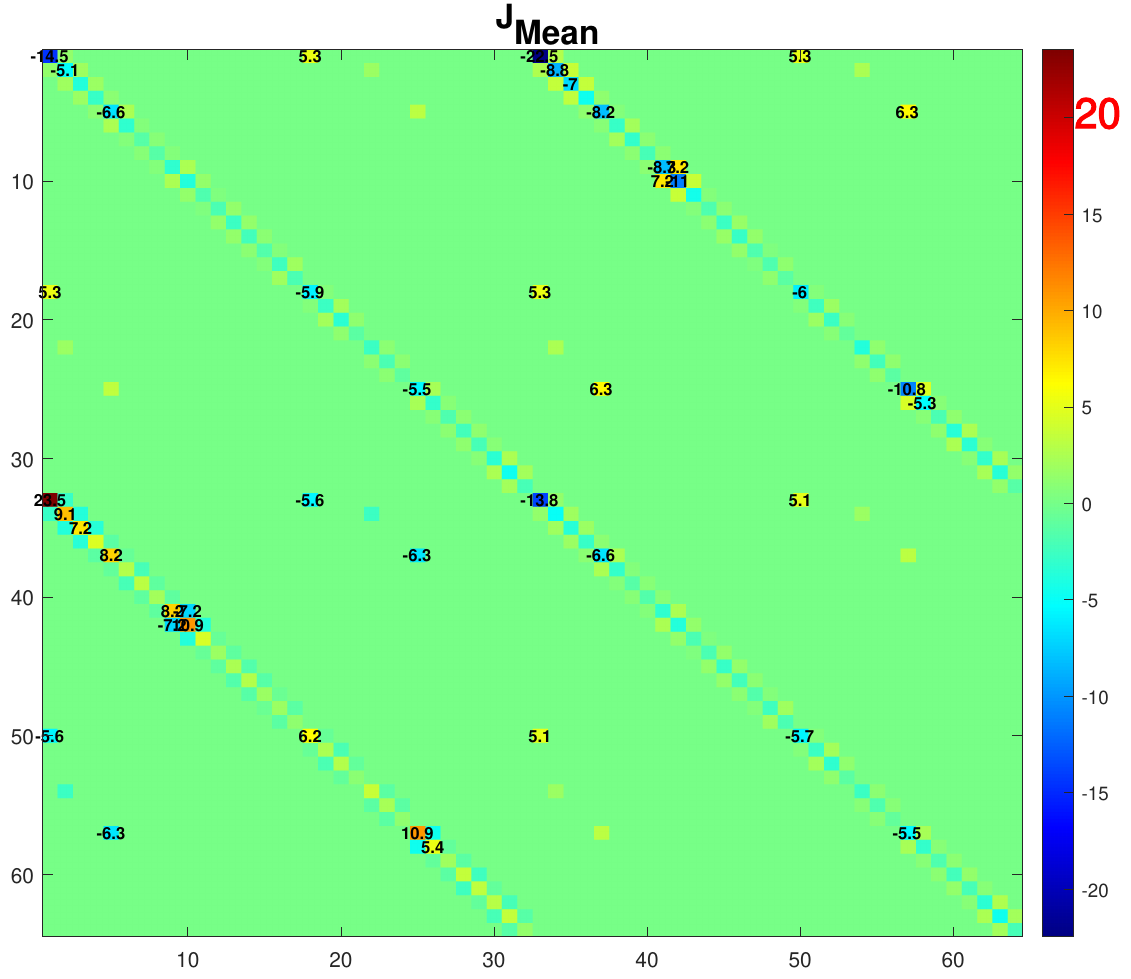}}
\subfloat[Standard Deviation :  $\mathbf J_{\text{SD}}$]{\label{fig:Jstd}
\includegraphics[width=0.24\textwidth]{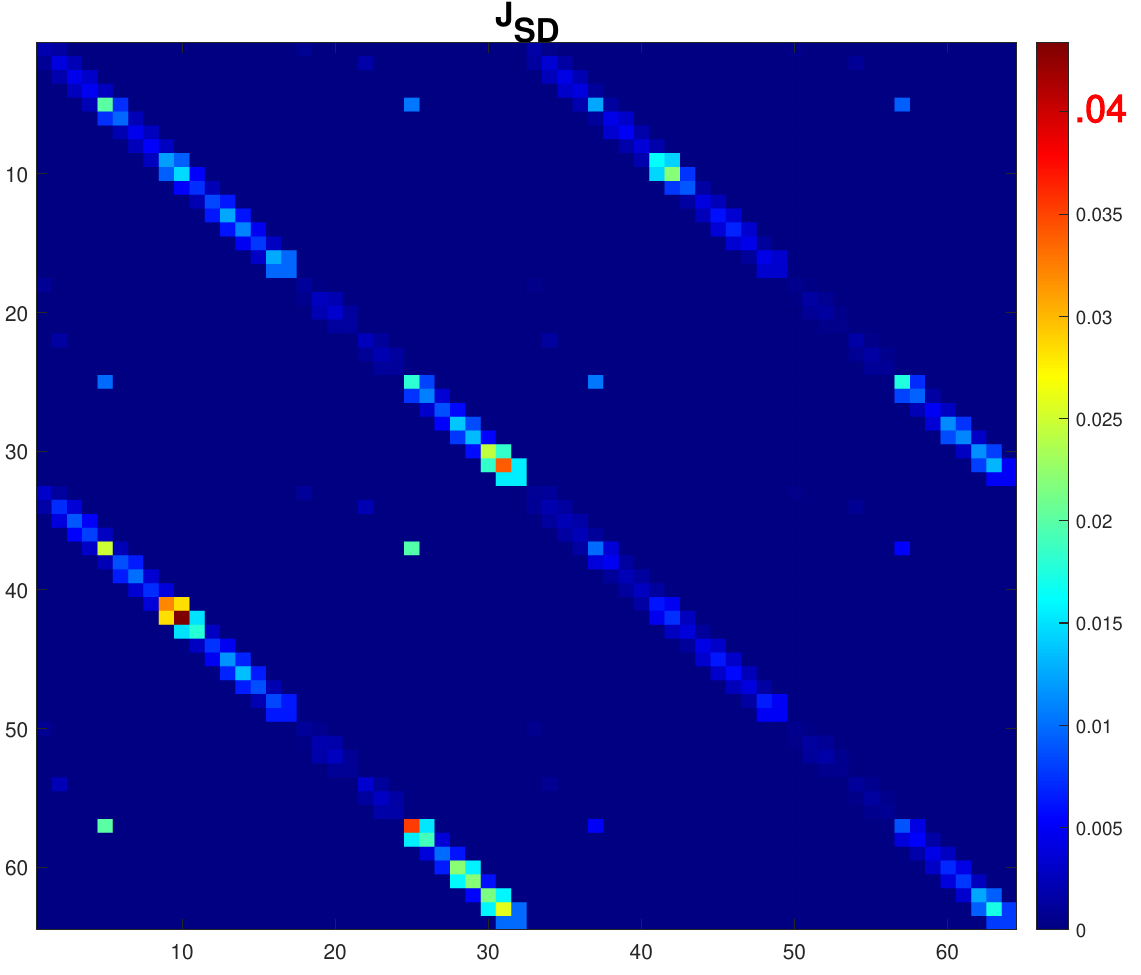}}
\caption{Basic  Statistical Information of $\mathbf J$ in Period $\text{T}_1$}
\label{fig:dailyJ0}
\end{figure}

In this case, we take the period $\text{T}_1$ ($ 1 \!\sim\! 720$) for study. The truth-value of $\mathbf J$ on each sampling point is calculated via  Eq.~\eqref{eq:HNKLend} in a model-based way.
The result validates that $\mathbf J$ \textbf{indeed keeps nearly constant} at around its mean $\mathbf J_{\text{Mean}}$ (Fig.~\ref{fig:Jmean}, 20 level), and with the standard deviation $\mathbf J_{\text{SD}}$ (Fig.~\ref{fig:Jstd}, 0.04 level). Therefore, it is reasonable to set $\mathbf J_{\text{Mean}}$ \textbf{as the benchmark} during this observation period $\text{T}_1$.

Defining $\Delta {{\mathbf{x}}^{\left( k \right)}}\!\triangleq\! {{\mathbf{x}}^{\left( k+1 \right)}}\!-\!{{\mathbf{x}}^{\left( k \right)}}$ and $\Delta {{\mathbf{y}}^{\left( k \right)}}\!\triangleq\! {{\mathbf{y}}^{\left( k+1 \right)}}\!-\!{{\mathbf{y}}^{\left( k \right)}}$, Eq. \eqref{eq:J1} is rewritten as $\Delta {{\mathbf{y}}^{\left( k \right)}}\!\approx\!\mathbf {J}^{\left( k \right)}\Delta {{\mathbf{x}}^{\left( k \right)}}$. \textbf{Since $\mathbf {J}$ keeps nearly constant} during $\text{T}_1$, the expression is reformulated as
\begin{equation}
\label{Eq:MMYX}
\mathbf B \!\approx\! \mathbf J\mathbf{A}
\end{equation}
where $\mathbf{J} \!\in\! {{\mathbb{R}}^{K\!\times\!K}} $, $\mathbf{B} \!=\! \left[ {\Delta {{{\mathbf{y}}^{(1)}}}, \cdots ,\Delta  {{{\mathbf{y}}^{(T)}}}} \right] \!\in\! {{\mathbb{R}}^{K\!\times\!T}}$, and $\mathbf{A}\! =\! \left[ {\Delta {{{\mathbf{x}}^{(1)}}}, \cdots ,\Delta  {{{\mathbf{x}}^{(T)}}}} \right] \!\in\! {{\mathbb{R}}^{K\!\times\!T}}$.

The least square method is {the first and most obvious choice} as the solution to the regression problem formulated as Eq.~\eqref{Eq:MMYX}.
It is capable of handling the scenarios where the network topologies $\mathbf Y$ are \textbf{unreliable or even totally unavailable}, and thus, $\mathbf Y$ are no longer essential information. This property agrees with our assumption in Sec.~\ref{Sec:Assumpt}.
Conversely, the result of $\mathbf {J}$ estimation inherently contains the most up-to-date information about $\mathbf Y$.

\begin{figure*}[ht]
\centering
\subfloat[\textbf{OLS}: without Error (0.07 level)]{\label{fig:OLSwithoutErrors}
\includegraphics[width=0.24\textwidth]{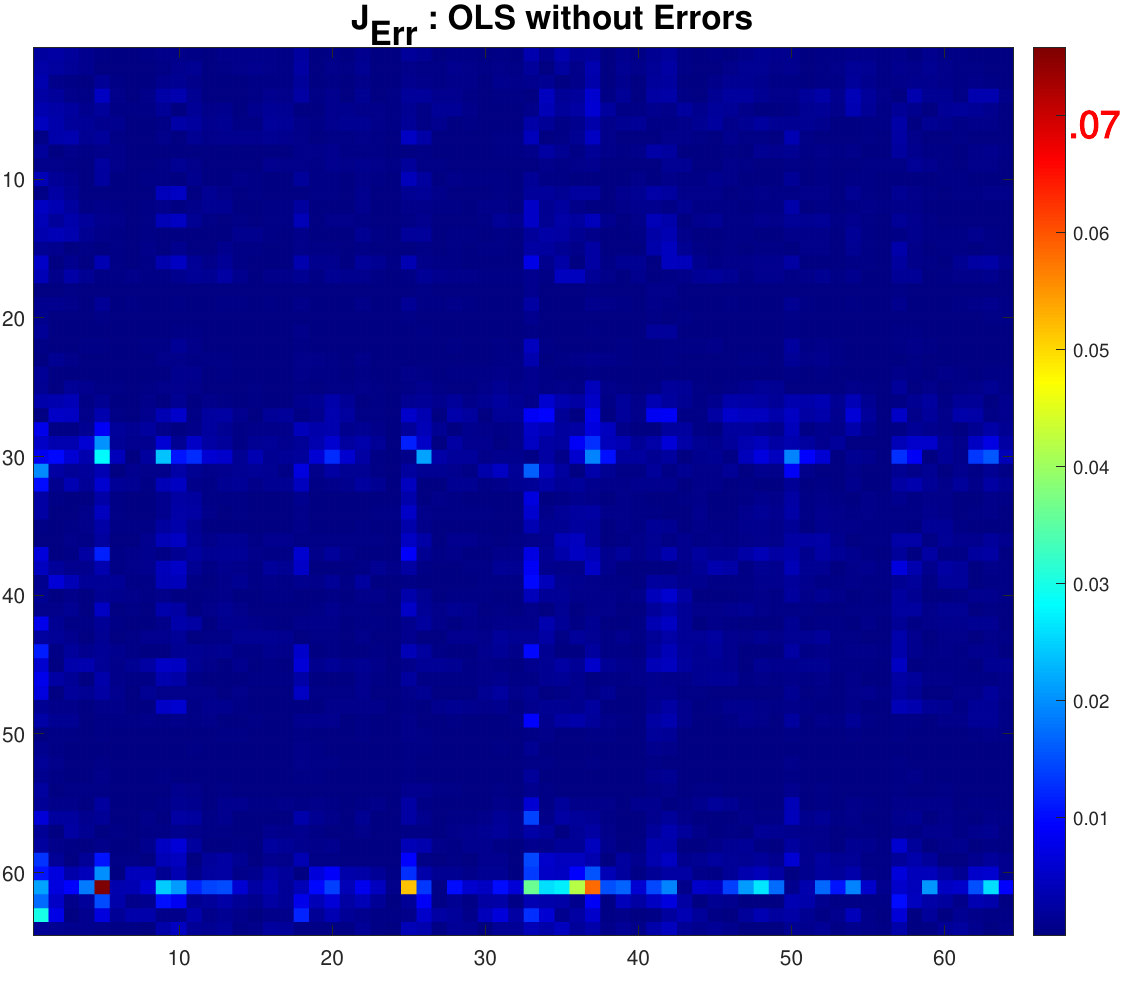}}
\subfloat[Gaussian Error from $\mathbf {y}$ (0.3 level)]{\label{fig:OLSwithErrorY}
\includegraphics[width=0.24\textwidth]{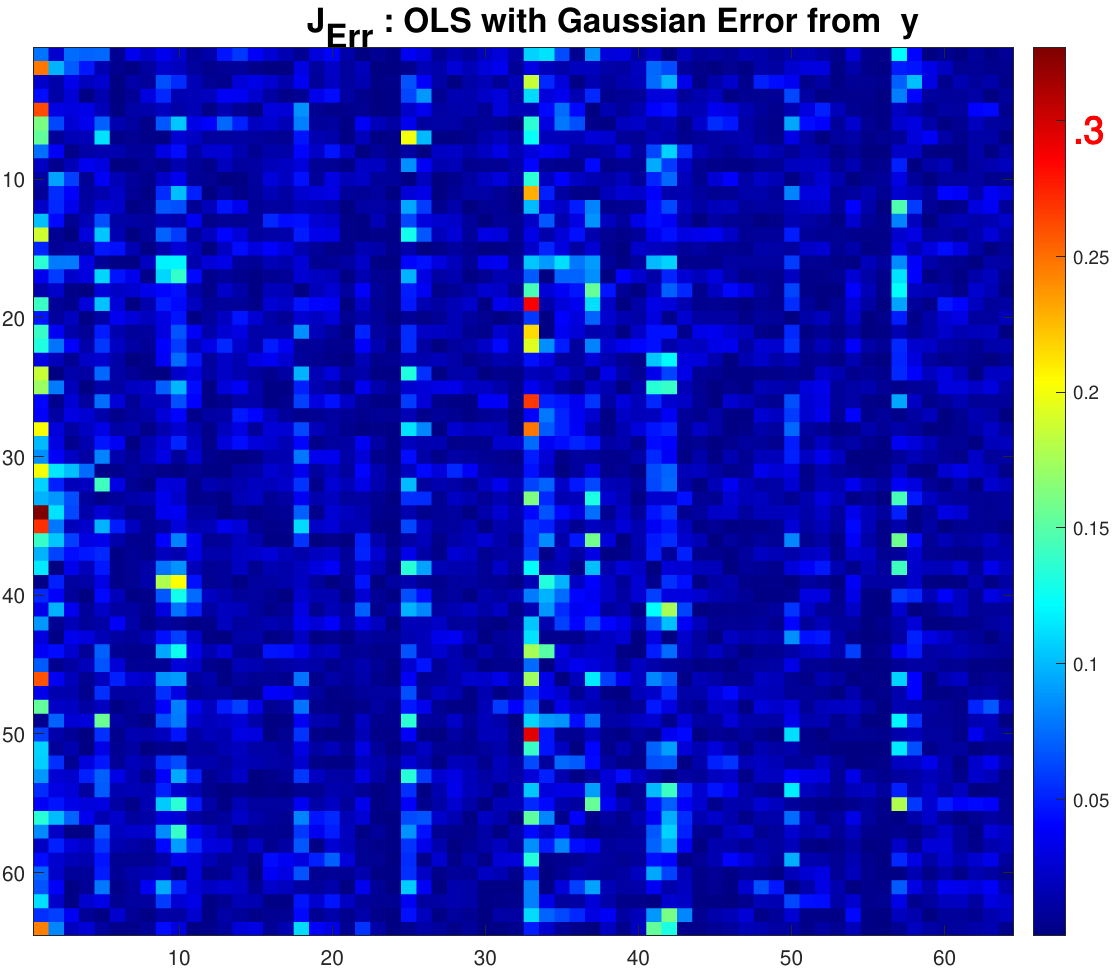}}
\subfloat[Error from both $\mathbf {y}$ and $\mathbf {x}$ (-)]{\label{fig:OLSwithErrorYX}
\includegraphics[width=0.24\textwidth]{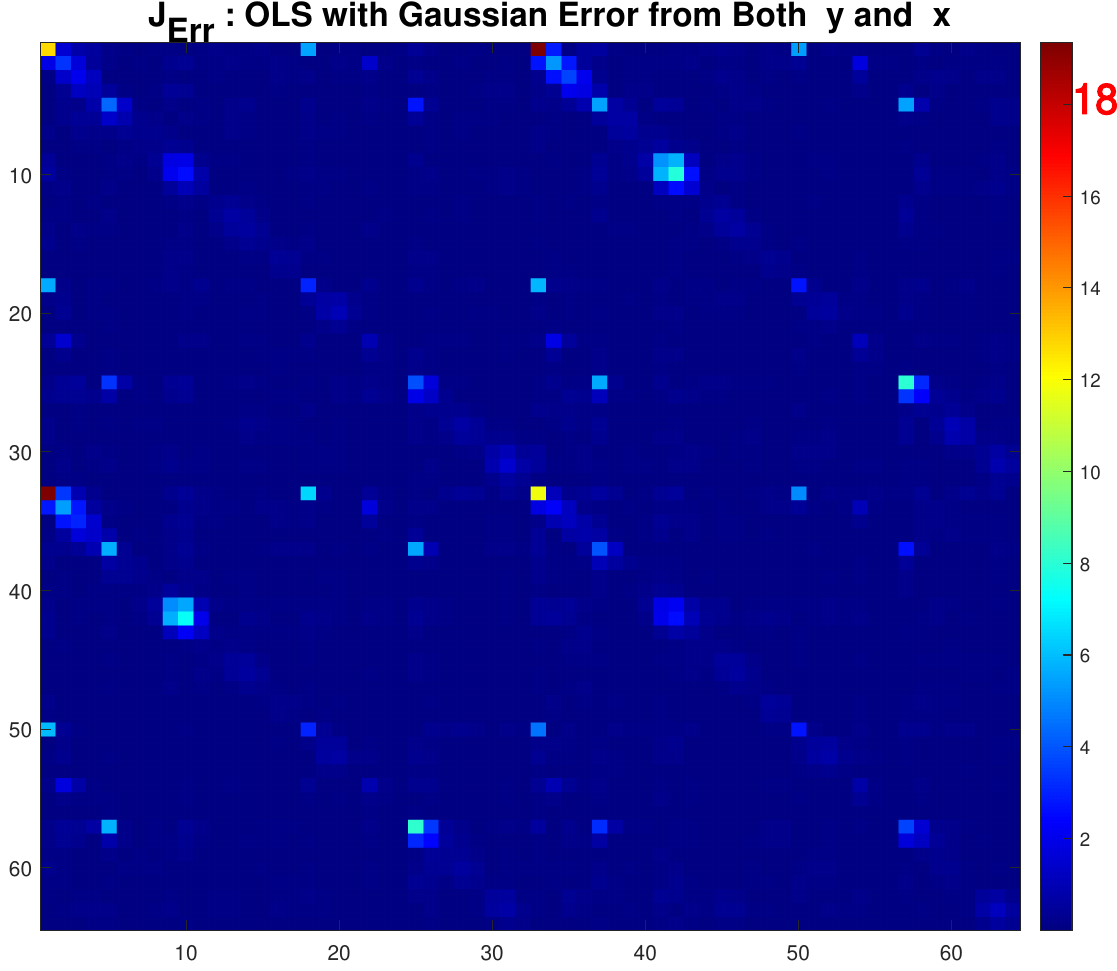}}
\subfloat[Non-Gaussian Error from both (-)]{\label{fig:OLSwithARErrorYX}
\includegraphics[width=0.24\textwidth]{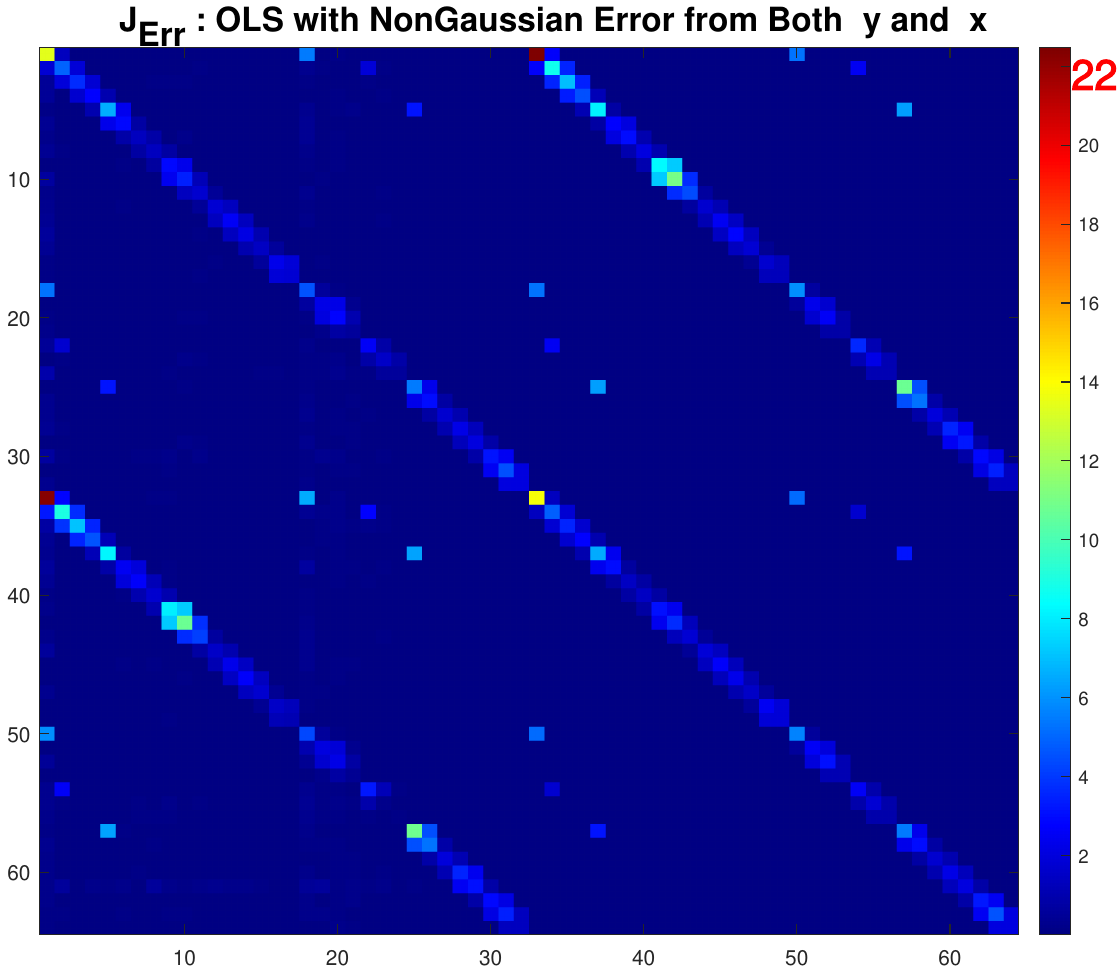}}

\subfloat[\textbf{TLS}: without Error (0.07 level)]{\label{fig:TLSwithoutErrors}
\includegraphics[width=0.24\textwidth]{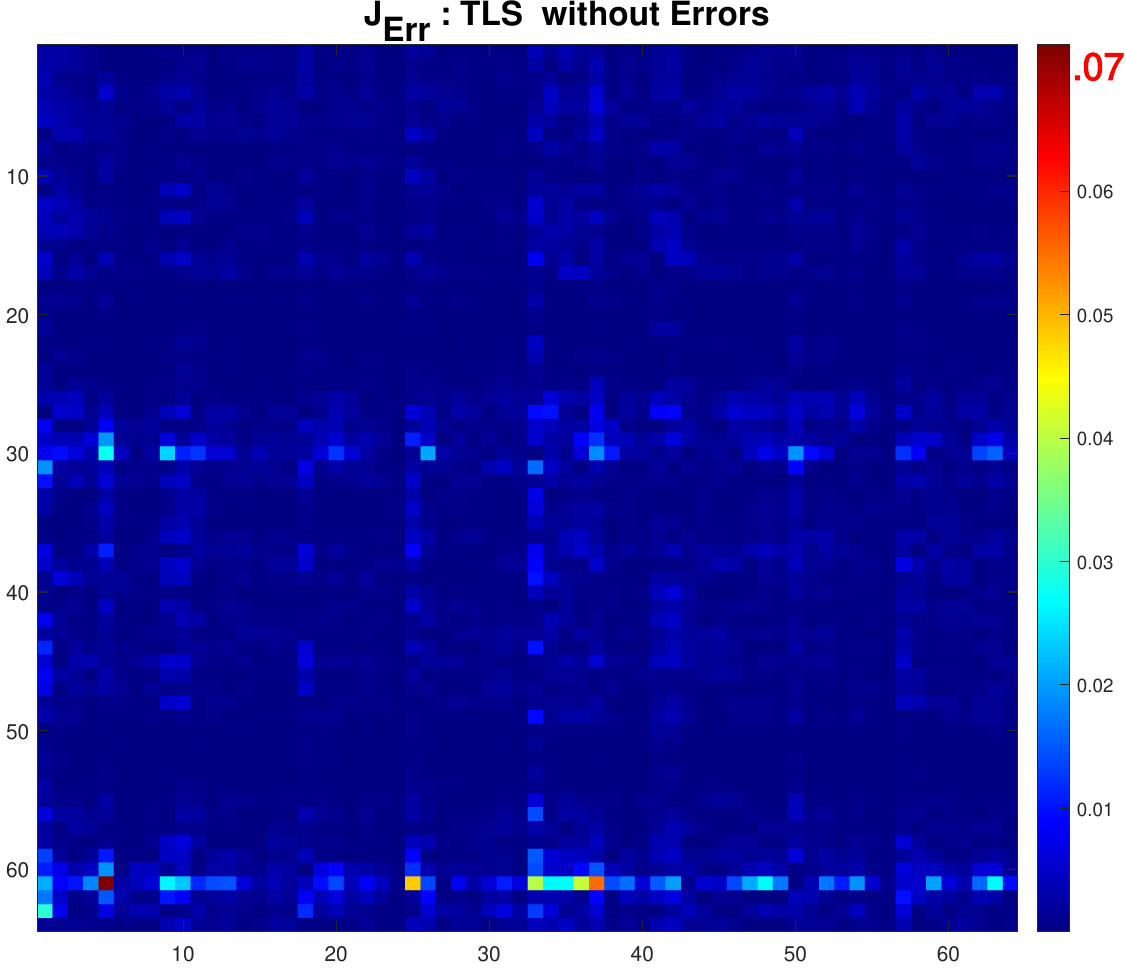}}
\subfloat[Gaussian Error from $\mathbf {y}$ (0.25 level)]{\label{fig:TLSwithErrorY}
\includegraphics[width=0.24\textwidth]{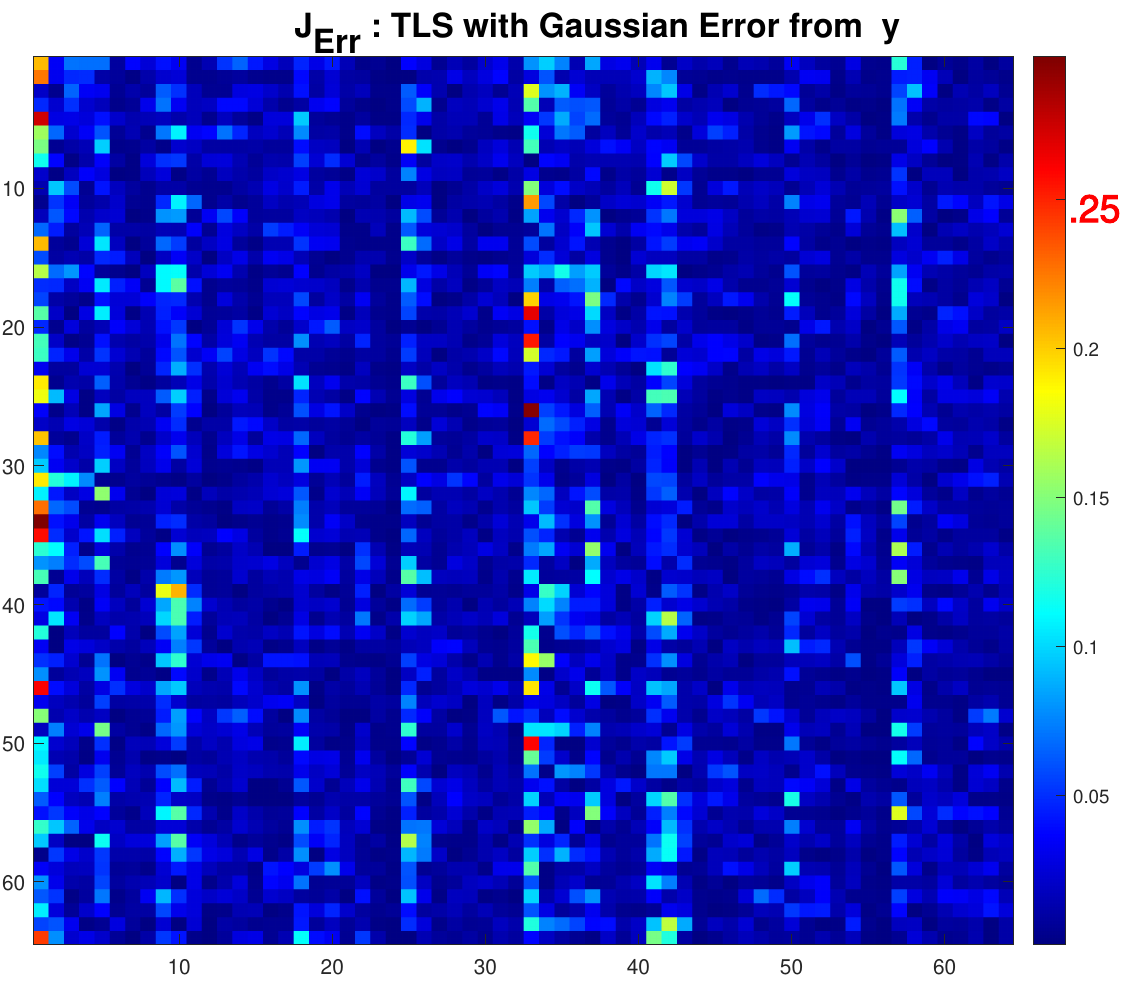}}
\subfloat[Error from both $\mathbf {y}$ and  $\mathbf {x}$ (4 level)]{\label{fig:TLSwithErrorYX}
\includegraphics[width=0.24\textwidth]{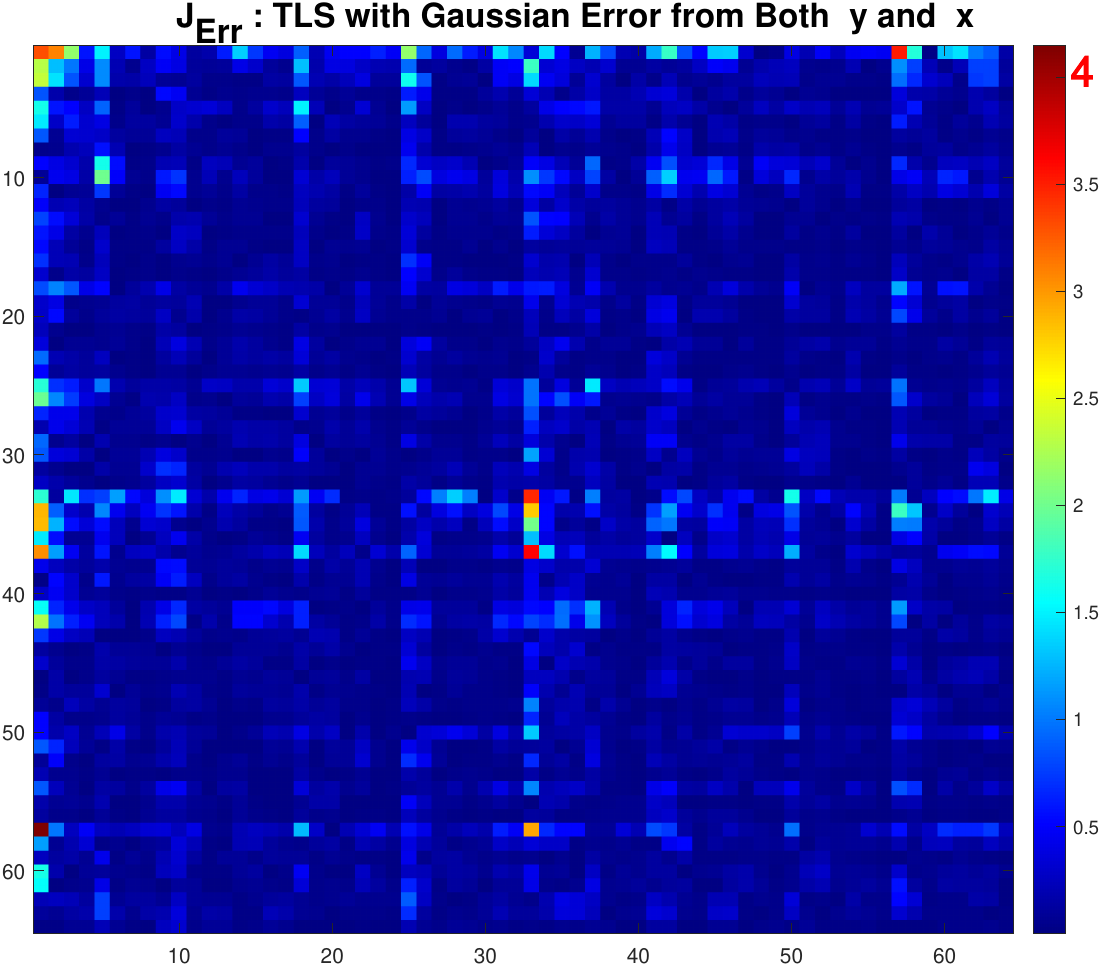}}
\subfloat[Non-Gaussian Error from both (-)]{\label{fig:TLSwithARErrorYX}
\includegraphics[width=0.24\textwidth]{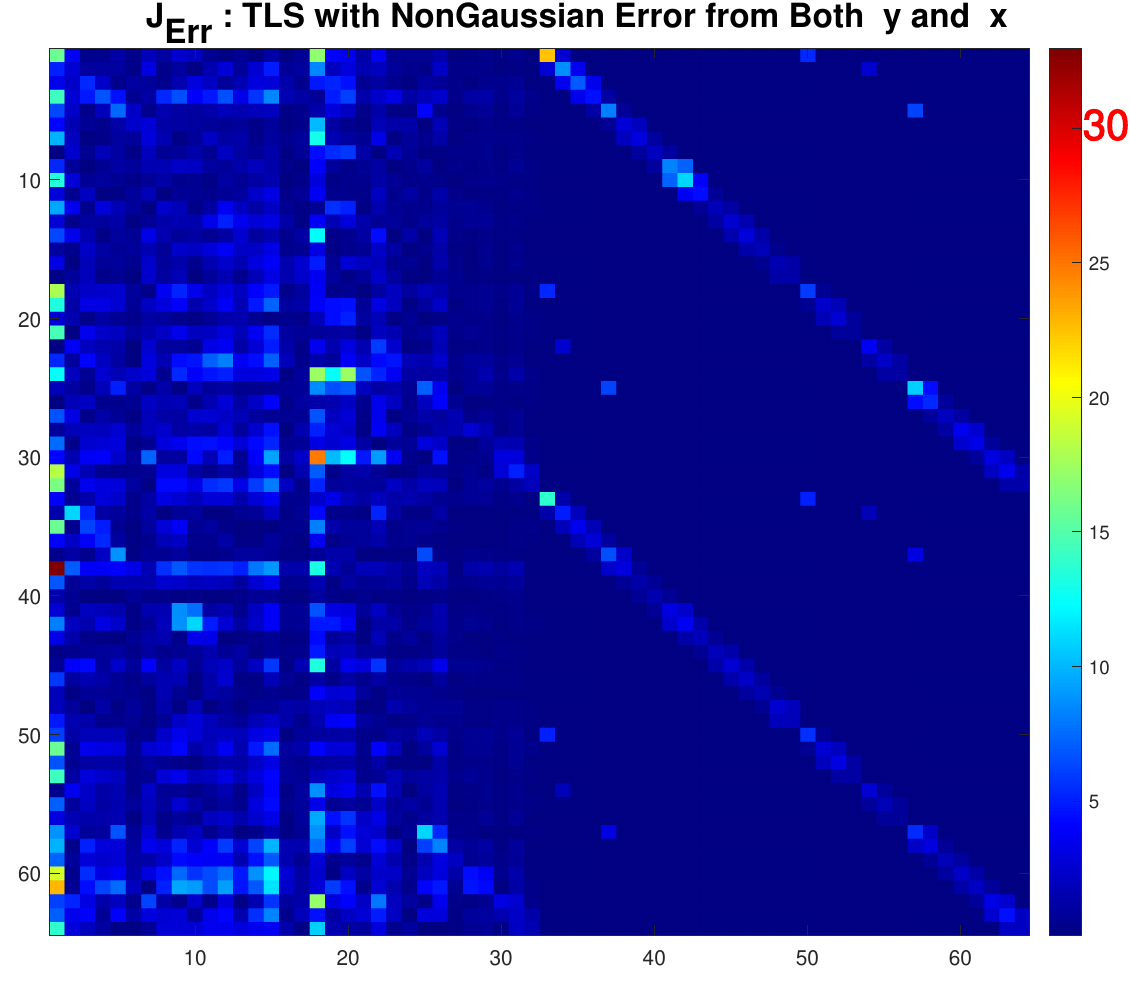}}

\caption{Performance of OLS and TLS on $\mathbf J$ Estimation with Different Types of Noise}
\label{fig:reusltJE}
\end{figure*}

In particular, ordinary least square (OLS)  and total least square (TLS)~\cite{Passerini2017Power} are tested, and numerous scenarios with different types of noise are studied.  Fig.~\ref{fig:reusltJE} shows the results.

\begin{enumerate}[1.]
\item Fig.~\ref{fig:OLSwithoutErrors} and \ref{fig:TLSwithoutErrors} tell that both OLS and TLS perform \textbf{well} ($\mathbf J_\text{Err}$ is at the same order as $\mathbf J_\text{SD}$; $\mathbf J_\text{Err}$---difference between the estimated values and the benchmark $\mathbf J_\text{Mean}$) \textbf{when there is no error} on neither $\mathbf{y}$ side ($\mathbf B$ in Eq~\ref{Eq:MMYX}) nor $\mathbf{x}$ side ($\mathbf A$).
\item Fig.~\ref{fig:OLSwithErrorY} and \ref{fig:TLSwithErrorY} tell that their performances reduce \textbf{from good level to acceptable level} when some Gaussian error (5\%) injects into $\mathbf{y}$ ($\mathbf x$ is assumed to be \textbf{error free}).
\item Fig.~\ref{fig:OLSwithErrorYX} and \ref{fig:TLSwithErrorYX} tell that when the Gaussian error (5\%) comes \textbf{from both $\mathbf{y}$ and $\mathbf{x}$}, TLS becomes the \textbf{only option} to reach a \textbf{barely-passing result}. TLS is a type of error-in-variables regression, a least squares data modeling technique in which observational error on both dependent and independent variables is taken into account~\cite{wiki2018TLS}.
\item However, if the noise \textbf{does not follow i.i.d. Gaussian distribution}, as the aforementioned renewables-derived uncertainties, both OLS and TLS  \textbf{fail} in this kind of regression task. These uncertainties, which are \textbf{analytically intractable} under conventional framework, will \textbf{almost certainly}  lead to bad results \textbf{without a proper treatment}, as illustrated in Fig.~\ref{fig:OLSwithARErrorYX} and \ref{fig:TLSwithARErrorYX}. This is the \textbf{primary motivation} for our proposed hybrid framework.
\end{enumerate}

\subsection{Elementary RMT-based Analysis}
\label{Sec:CaseERMT}
To make these renewables-derived uncertainties analytically tractable, we have to study the problem in a high-dimensional space.
Under the RMT framework provided in our previous work \cite{he2015arch}, we gain insight the uncertainties from the spectrum aspect via high-dimensional analysis.

\begin{figure}[htbp]
\centering
\subfloat[ESD of $\mathbf C_{1\_1}$: Model 1 in $\text{T}_1$]{\label{fig:X1T1}
\includegraphics[width=0.22\textwidth]{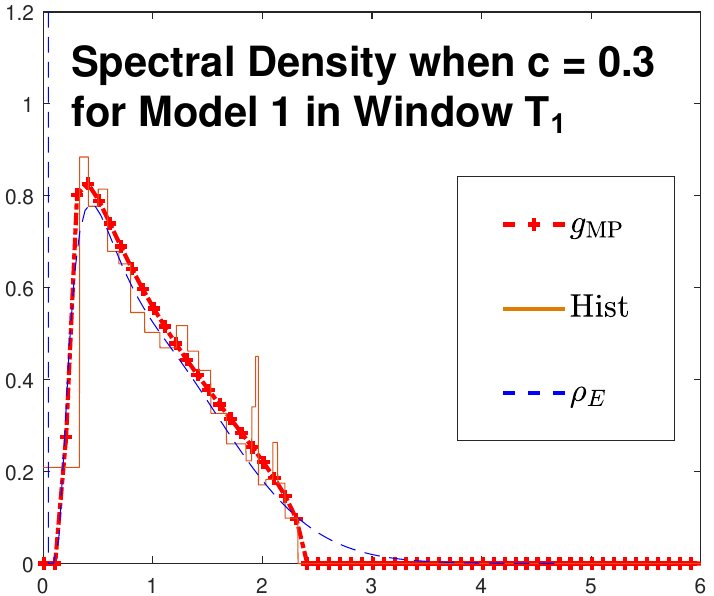}}
\subfloat[ESD of $\mathbf C_{1\_2}$: Model 1 in $\text{T}_2$]{\label{fig:X1T2}
\includegraphics[width=0.22\textwidth]{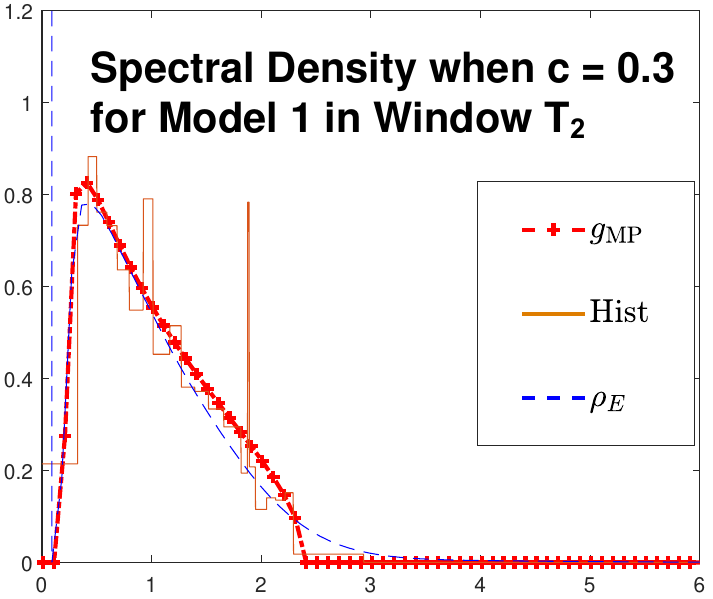}}

\subfloat[ESD of $\mathbf C_{1\_5}$: Model 1 in $\text{T}_5$]{\label{fig:X1T5}
\includegraphics[width=0.22\textwidth]{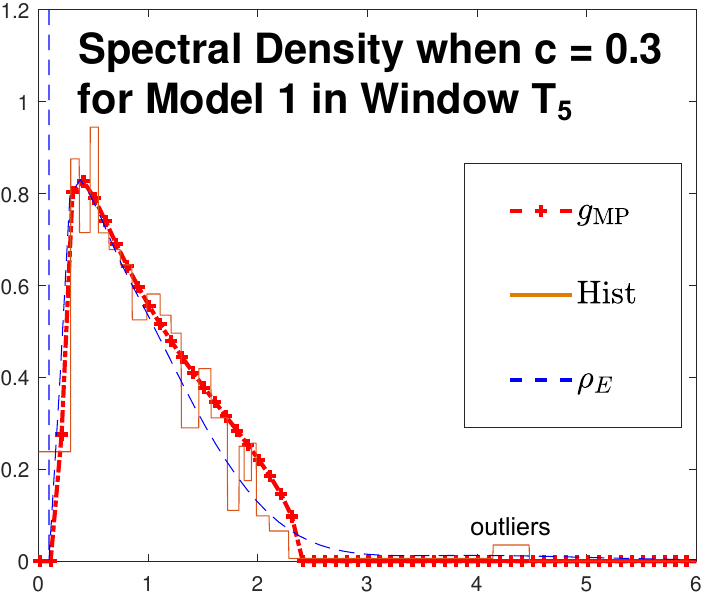}}
\subfloat[ESD of $\mathbf C_{3\_1}$ with Different $p$]{\label{fig:X3T1}
\includegraphics[width=0.22\textwidth]{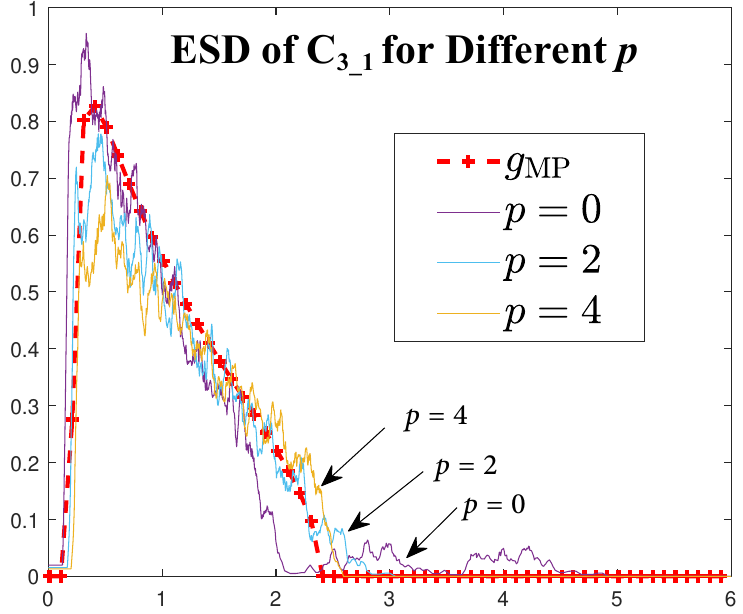}}
\caption{ESD of $\mathbf C_{m\_j}$ for Model $m$ in Period $\text{T}_j$}
\label{fig:ESDX1X3}
\end{figure}

Fig.~\ref{fig:ESDX1X3} depicts the analysis result for Model $\text M_m$ in Period $\text{T}_j$.
The `$\bm g_\text{T}$' Curve is the theoretical M-P Law spectral density as given in Eq.~\eqref{eq:D2}.
The `Hist' Curve means histogram for the ESD.
%First, we set the number of factors $p\!=\!1$ in Eq.~\eqref{eq:FactM}  by removing the major system inherent part referring to \cite{yeo2016random}, and thus convert difference $\mathbf{V}$ into residues $\mathbf{R}_1$.
First, we set factor numbers $p$ in Eq.~\eqref{eq:FactM} to convert difference $\mathbf{V}$ into residues $\mathbf{R}$.
Then we calculated the ESD of $\mathbf C_{m\_j}\!=\! \frac{1}{T}\mathbf R \mathbf R^{\text T}$ according to Eq.~\eqref{eq:ESDdef}.
The `$\bm \rho_\text{E}$' Curve is the probability density estimate of the `Hist' Curve using Kernel Smoothing Function (code `ksdensity($\cdot$)' in Matlab, for Model $\text M_1$) or Moving Average Function (code `smooth($\cdot$)', for $\text M_3$).

The metric space designed in  Sec. \ref{Sec:MetricDesign} enables us to \textbf{quantify} the TI performance of each bank model in spectrum space. The outliers tend to big and evident as the corresponding model becomes deviant,  and the deviation will lead to a large $d(\mathbf{V}_\text{ob}, \hat{\mathbf{V}}_{m\_j})\!=\!|\mathbf{V}_{m\_j}|_{\mathcal D}$ as defined in Eq. \eqref{eq:Matric}.
%This kind of spectrum metric space based indicators depend on the temporal-spatial data block, and therefore they are robust against partial data error, missing and unsynchronized. The details can be found in our previous work  \cite{he2016les}.

\subsection{FA Analysis and Time-Series Analysis}
%Sec. \ref{Sec:Resi} tells that FA provides a way to decompose the data matrix into systematic information part and idiosyncratic noise part.
For each difference-derived random matrix, e.g. $\mathbf V_{3\_1}$, we calculate its ESD with a different factor numbers $p$, and then obtain the results as shown in Fig.~\ref{fig:X3T1}.
As we increase factor numbers $p$, the outliers are alleviated. This phenomenon agrees with the fact that FA is often used for dimension reduction in sampling data with underlying constructs, i.e. converting $\mathbf V_{m\_j}$ into ${{\mathbf{L}}^{\left( p \right)}}{{\mathbf{F}}^{\left( p \right)}}$ following Eq.~\eqref{eq:FactM}.
However, the residues part $\mathbf R_{m\_j}$ \textbf{could also have some latent construct}. For instance, the randomness caused by a wind following AR model with coefficients $b$.
This statistic property \textbf{cannot be eliminated} simply by increasing $p.$ Fortunately, Ref.~\cite{burda2010random}  applies RMT to derive spectral density of large sample covariance matrices generated by multivariate ARMA processes \textbf{in analytic forms} (Eq.~\ref{eq:crossautocorrelated}$\rightarrow$\ref{Eq:polynomial}$\rightarrow$\ref{Eq:MzGz}$\rightarrow$\ref{Eq:inverse_green_function}). Following Ref.~\cite{burda2010random}, we push forwards our research on the residues $\mathbf R_{m\_j}.$

\begin{figure}[htbp]
\centering
\includegraphics[width=0.46\textwidth]{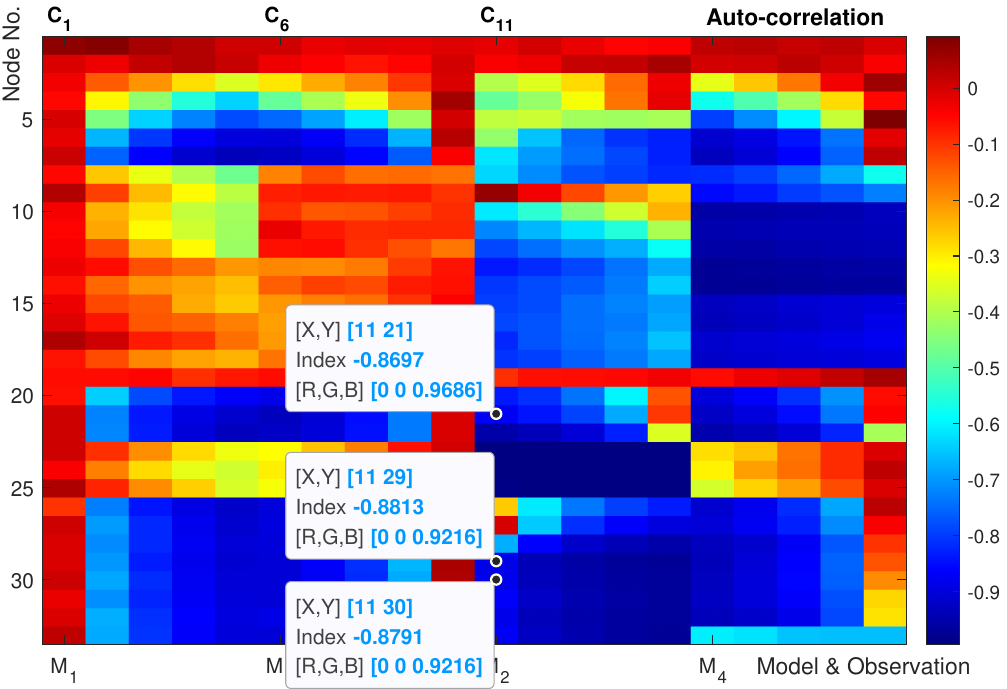}
\caption{Estimated auto-correlation coefficient $\hat b$ of $\mathbf V_{m\_j}$}
\label{fig:autocorrelation}
\end{figure}

Temporal analysis is conducted first by estimating the auto-correlation coefficient $b$  of $\mathbf R_{m\_j}$ using Burg's method (code `arburg($\cdot$)' in Matlab). If the picked model perfectly matches the real grid, the renewables-derived auto-correlation would be eliminated, and only (Gaussian) measurement error remains. Fig.~\ref{fig:autocorrelation} validates this---all the node on $\mathbf V_{1\_1}$ (Column $\text C_1$) and $\mathbf V_{2\_5}$ ($\text C_{10}$) are of small auto-correlation  ($\hat b\approx 0$), and therefore we should \textbf{accept the hypothesis} that Model $\text M_1$ matches the real system in Period $\text{T}_1$, and $\text M_2$ in Period $\text{T}_5$.

Fig.~\ref{fig:X1T1} also depicts the phenomenon that only measurement error remains---$\mathbf V_{1\_1}$-derived ESD does closely match the theoretical `$\bm g_\text{T}$' Curve (M-P Law)  and no obvious outliers exist.
%Besides, we can find that for a certain model, $\hat b$ has a gradual change with our observation window. This is an important benefit of a well designing metric space.
Besides, we can find that the values of the nodes close to the reference bus (e.g. Node 2, 3, 19) are usually stable around 0. The phenomenon that these nodes are insusceptible to renewables is  consistent with our common sense.

\subsection{Jointly Temporal-spatial Analysis with Latent Structure}
M-P Law can nicely model $\mathbf R_{1\_1}$ in some sense. Then some open questions are raised, for example: 1) How to model other columns, e.g., Column 11 ($\mathbf R_{3\_1}$)? 2) Can we extract some information from them, and how? To address these questions, jointly temporal-spatial analysis is discussed.

We revisit our prior information to find out the causes which may decide/influence the statistical properties of $\mathbf R_{m\_j}$.
One major cause is the two independent renewables on Node 20 and Node 31.  From the local field data we know that their power outputs follow AR process with some latent structure.
Another major cause is the inherent topology $\mathbf Y$, although it is unknown and may have a transformation at some time point.

Then we conduct the analysis with the \textbf{data from a few nodes} but not all of them. This is practical when the advanced sensors such as $\mu$PMUs are only deployed on some important buses.
RMT-framework \textbf{inherently supports} statistical analysis with \textbf{data only from a subset of nodes}---the data matrix can be naturally divided into data blocks \textbf{without additional error}, but this is not true for mechanism models. Our previous work~\cite{he2016les} gives a discussion on this RMT-framework property.

\begin{figure}[htbp]
\centering
\includegraphics[width=0.48\textwidth]{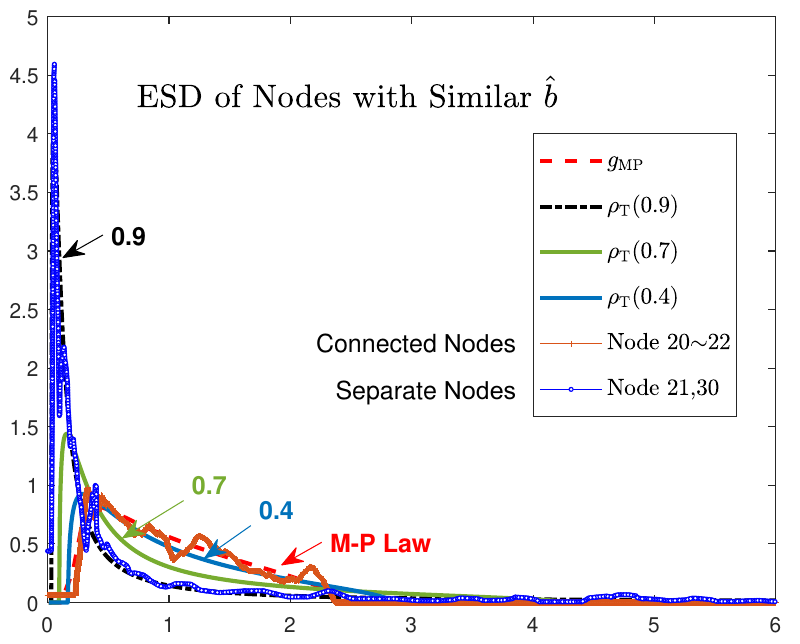}
\caption{Jointly Spatial-temporal Analysis to Grid Nodes}
\label{fig:STAna}
\end{figure}

For Column 11 ($\mathbf R_{3\_1}$), we take the renewables-influenced nodes' data ($b\approx 0.9$) into account, and then make a \textbf{jointly temporal-space analysis} following Sec.~\ref{Sec:Resi}. The coefficients $\hat b$ of these influenced nodes are similar.
With the prior knowledge of Model $\text M_3$ stored in the bank, we divide these influenced nodes into three parts: 1) Node 6, 7; 2) Node 20$\sim$22; and 3) Node 29$\sim$33.
Then we study their \textbf{cross-correlation} under this division---the closely connected nodes \textbf{must show strong correlation}, while the separated nodes show the independence.  Based on this property, we use the theoretical spectral density  $\bm \rho _{\text{T}}(b)$ to test them, and the results are given in Fig.~\ref{fig:STAna}.

The ESD of relevant data derived from separated nodes \textbf{closely matches the theoretical density} $\bm \rho _{\text{T}}(0.9)$. This phenomenon is built upon the premise of
the Assumption \uppercase\expandafter{\romannumeral 1}  in Sec.~\ref{sec:crosscorr}, i.e.  $\hat{\mathbf R}$ has sufficiently negligible cross-correlation: cross-covariances matrix $\mathbf A_N \!\approx\! \mathbf I_{N\times N}$---the randomness component of these separated nodes are influenced by renewables with \textbf{independent behaviors}. This independence is often reasonable especially for an integrated energy system (IES) with \textbf{diverse sources}. While for those closely connected nodes (Node 20$\sim$22 in this case), the \textbf{independence condition is violated}, so there is no consistency between $\bm \rho_{\text E}$ and $\bm \rho_{\text T}(0.9)$.

\subsection{Test with IEEE 85-bus Network}
\begin{figure}[htbp]
\centering
\includegraphics[width=0.46\textwidth]{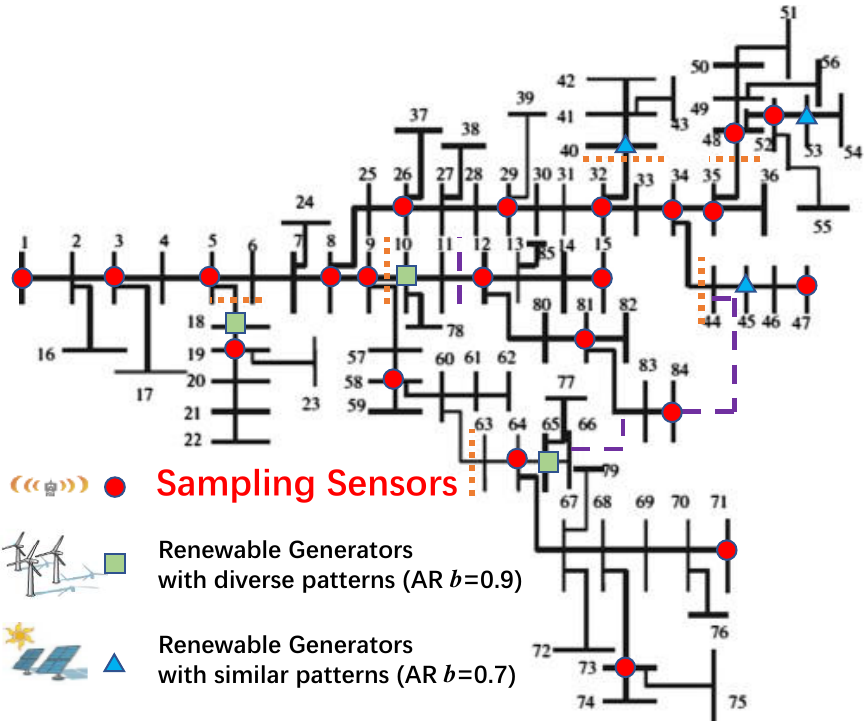}
\caption{IEEE 85-bus radial distribution systems}
\label{fig:IEEE85}
\end{figure}

In addition, we test our framework using IEEE 85-bus radial distribution systems. The sampling sensors, renewable generators with diverse/similar patterns are deployed as Fig.~\ref{fig:IEEE85}.

\begin{figure}[htbp]
\centering
\subfloat[Estimated auto-correlation coefficient $\hat b$ of $\mathbf V_{m}$]{\label{fig:Ar_b85}
\includegraphics[width=0.48\textwidth]{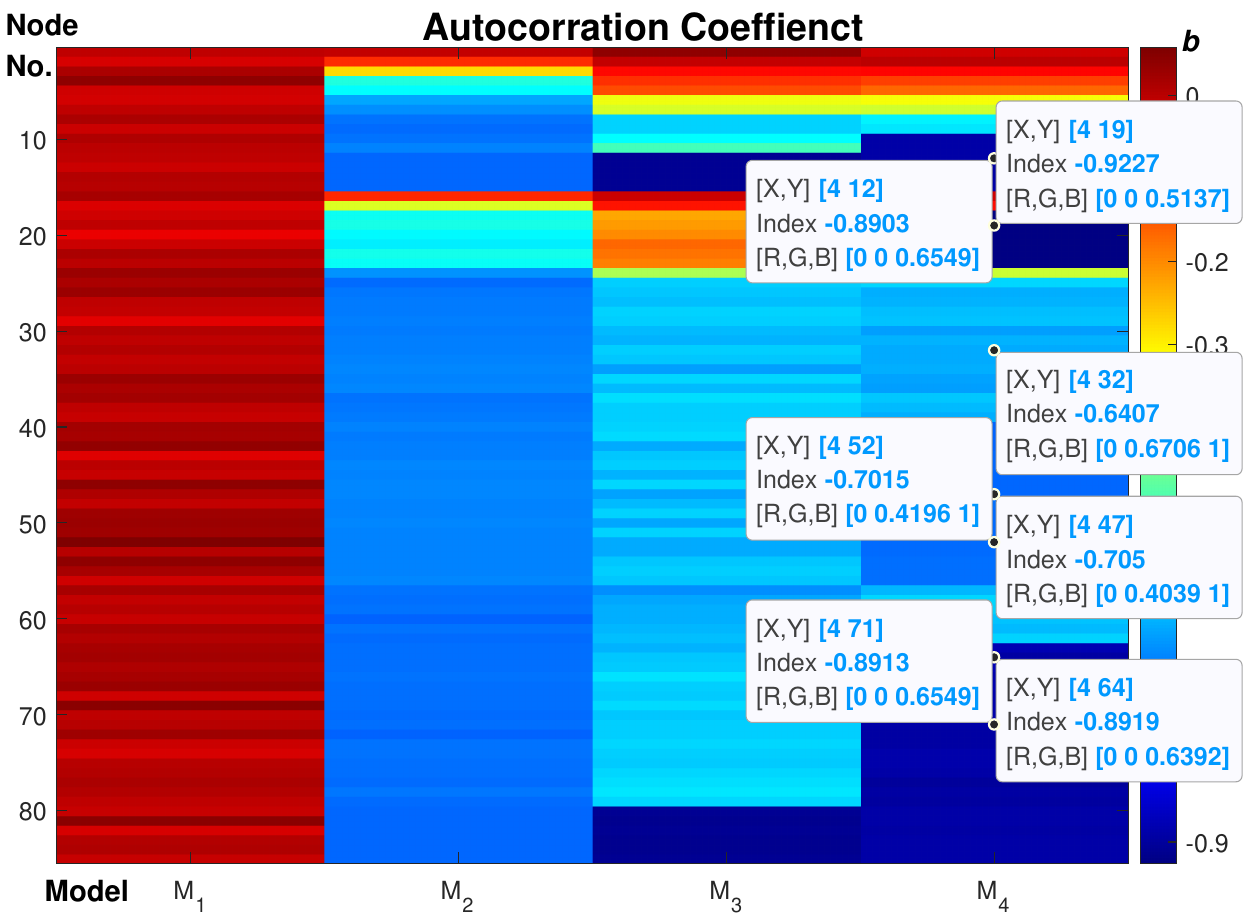}}

\subfloat[Jointly Spatial-temporal Analysis ]{\label{fig:JSA85}
\includegraphics[width=0.48\textwidth]{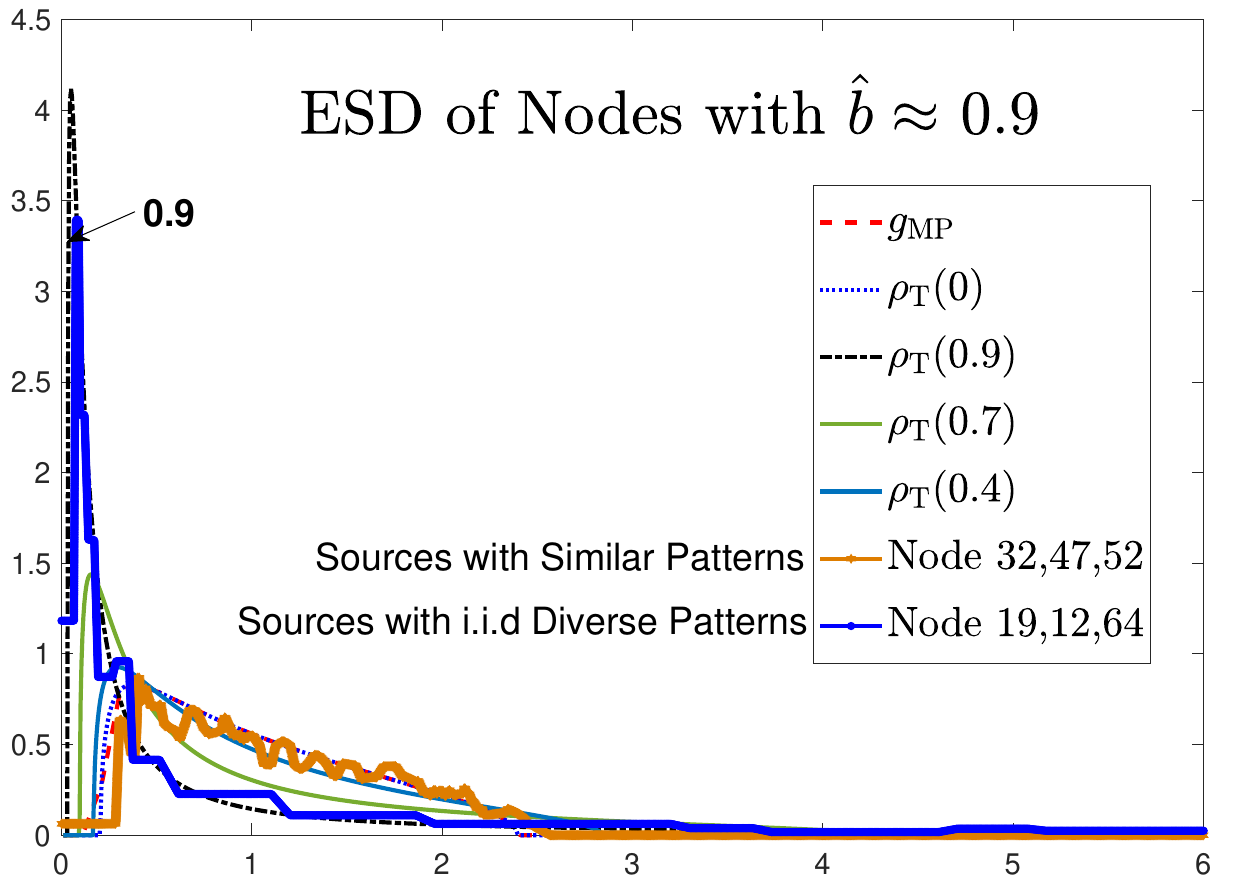}}
\caption{Result of  IEEE 85-bus Radial Distribution Systems}
\label{fig:Res85}
\end{figure}

As a distribution grid usually operates in open loop, we just test the \textbf{pair switch} of the normally closed branches and the normally open branches. In particular, we test the pair switch of the normally closed branches $\text B_{11-12}$ (closed$\rightarrow$open) companied with the normally open branch $\text B_{44-84}$ (open$\rightarrow$closed) in Model $\text M_{2}$, and with $\text B_{66-83}$ (open$\rightarrow$closed) in $\text M_{3}$, respectively. \textbf{The error of branch impedance} is also tested---$\text M_{4}$ tests $\text B_{5-18}$, $\text B_{9-10}$, $\text B_{60-63}$, $\text B_{32-40}$, and $\text B_{35-48}$.
Similar to Fig.~\ref{fig:autocorrelation} and \ref{fig:STAna}, Fig.~\ref{fig:Ar_b85} shows the time-series information, and Fig.~\ref{fig:JSA85} shows the jointly spatial-temporal analysis results.

The results in Fig.~\ref{fig:Res85} validate the hybrid framework again. This framework is suitable to the scenarios \textbf{when the renewables-behavior dominates our observed data}.
For the nodes influenced by multiple sources, such as Node $25\!\sim \!32,$ however, it is hard to model them in practice.
Under an ideal scenario, independent component analysis (ICA) or free component analysis (FCA) \cite{wu2019free} may be applied to separate the mixed signal into additive (independent)subcomponents. The combination of ICA/FCA and our framework offers potential for a more complex scenario.

%
%
%
%\subsubsection{RMT-based Analytics for Residues}
%{\text{\\}}
%
%For the estimation bias matrix, we conduct RMT-based analytics with factor models as mentioned in Section \ref{Sec:Resi}, and the analytics are obtained as shown in Fig.~\ref{fig:fm1} and \ref{fig:fm3}. `Estimation Bias 3', which gets a much better performance than `Estimation Bias 1', has a similar statistical trend curve but much fewer outliers. And from
%Fig.~\ref{fig:bb1}, we knows that  for `Estimation 1' there exist some duplications in the branch description file.
%This phenomenon indicates that the estimation result is sensitive to up-to-date topology parameters.

\section{Conclusion}
\label{section: concl}
\normalsize{}
This paper explores several high-dimensional analytics in the context of topology identification. We propose a hybrid framework, by tying AR model, FA, and RMT together, to handle the renewables-derived uncertainties in the form of multiple time-series.
Our framework, through \textbf{a systematic and theoretical processing}, makes these uncertainties analytically tractable, and \textbf{is immune to fixed measurement error}.

Several future studies are in order. Clearly, further research is needed to employ the \textbf{more general residue modeling}, for which we can calculate the spectral density readily. For example, as described in \cite{burda2010random}, if considering vector ARMA(1,1) processes, we have up to 6th-order polynomial equations.
Obviously, compared to i.i.d. Gaussian noise, the joint temporal model (AR) and spatial model (FA) oftentimes provide more \textbf{flexible and rigours} models and analyses on renewables-derived uncertainties.
Besides, the framework is capable to handle comprehensive behavior (on the nodes influenced by multiple sources) with the help of existing algorithm such as ICA.
The {combination} of {conventional tasks in  power system} with {novel tools in data science} is a long-term goal in our community, especially in big data era.
In addition, this hybrid framework can be extended to an integrated energy system, in which randomness and independence is more evident.

\bibliographystyle{IEEEtran}
\bibliography{helx}

\normalsize{}
\end{document}